\documentclass[fleqn,usenatbib]{mnras}

\usepackage{newtxtext,newtxmath}
\usepackage{rotating}

\usepackage[T1]{fontenc}
\usepackage{ae,aecompl}

\usepackage{graphicx}	
\usepackage{amsmath}	
\usepackage{amssymb}	
\newcommand{\angstrom}{\textup{\AA}}

\title[quasar chemical abundance]{The Evolution of Chemical Abundance in Quasar Broad Line Region}

\author[F. Xu et al.]{
Fei Xu,$^{1,2}$\thanks{E-mail: astroxf@163.com}
Fuyan Bian,$^{2,3,7}$\thanks{E-mail: fbian@eso.org}
Yue Shen,$^{4}$\
Wenwen Zuo,$^{5}$\
Xiaohui Fan,$^{6}$\
Zonghong Zhu$^{1}$\
\\
$^{1}$Department of Astronomy, Beijing Normal University\\
$^{2}$Research School of Astronomy \& Astrophysics, Mt Stromlo Observatory, Australian National University, Canberra, ACT 2611, Australia \\
$^{3}$European Southern Observatory, Alonso de C`ordova 3107, Casilla 19001, Vitacura, Santiago 19, Chile\\
$^{4}$Department of Astronomy, University of Illinois at Urbana-Champaign\\
$^{5}$Shanghai Astronomical Observatory, Chinese Academy of Sciences\\
$^{6}$Steward Observatory, University of Arizona\\
$^{7}$Stromlo Fellow
}

\date{Accepted 2nd July 2018. Received YYY; in original form ZZZ}

\pubyear{2015}

\begin{document}
\label{firstpage}
\pagerange{\pageref{firstpage}--\pageref{lastpage}}
\maketitle

\begin{abstract}
We study the relation between the metallicity of quasar broad line region (BLR) and black hole (BH) mass ($\rm 10^{7.5}M_{\sun} \sim 10^{10}M_{\sun}$) and quasar bolometric luminosity ($\rm 10^{44.6}erg/s \sim 10^{48} erg/s$) using a sample of {$\sim$130,000 quasars at $2.25\leq z\leq 5.25$ from Sloan Digital Sky Survey (SDSS) Data Release 12 (DR12)}. We generate composite spectra by stacking individual spectra in the same BH mass (bolometric luminosity) and redshift bins {and then estimate} the metallicity of quasar BLR using metallicity-sensitive broad emission-line flux ratios based on the photoionization models. We find a significant correlation between quasar BLR metallicity and BH mass (bolometric luminosity) but no correlation between quasar BLR metallicity and redshift. We also compare the metallicity of quasar BLR and that of host galaxies inferred from the mass-metallicity relation of star-forming galaxy at $z\sim2.3$ and $z\sim3.5$. We find quasar BLR metallicity is {0.3 $\sim$ 1.0 dex} higher than their host galaxies. This discrepancy cannot be interpreted by the uncertainty due to different metallicity diagnostic methods, mass-metallicity relation of galaxy, metallicity gradient in quasar host galaxies, BH mass estimation, the effect of different spectral energy distribution (SED) models, {and a few other potential sources of uncertainties.} We propose {a possibility} that the high metallicity in quasar BLR {might be} caused by metal enrichment from massive star formation {in the nucleus region of quasars or even the accretion disk.}
\end{abstract}




\section{Introduction}
Quasars are the most luminous subclass of active galactic nuclei (AGNs). They provide powerful tools to study the re-ionization process \citep{2006ARA&A..44..415F,2006AJ....132..117F}, supermassive BH (SMBH) growth \citep{2011Natur.474..616M,2015Natur.518..512W}, and chemical enrichment history  \citep{2002ApJ...564..592H, 2003ApJ...589..722D, 2006A&A...447..157N} at the early epoch of the Universe.  
Quasars are powered by accreting material onto the central SMBHs. The BLRs contain gas clouds close to the SMBHs, which are photoionized by the radiation field from the accretion disk of the SMBH. The ultraviolet and optical emission lines from BLRs are widely used to estimate the SMBH mass \citep{2006ApJ...641..689V, 2013BASI...41...61S, 2015ApJ...799..189Z} and the chemical abundance close to the SMBH \citep{2002ApJ...564..592H, 2006A&A...447..157N,2007AJ....134.1150J,2009A&A...494L..25J,2011A&A...527A.100M,2015Ap&SS.356..339M}.

Studies on the chemical abundance in the SMBH BLR provide insights on the chemical evolution of their host galaxies and shed light on the co-evolution of galaxies and their central SMBHs. The gas-phase metallicity in the BLR can be measured from broad emission-line flux ratios. Robust metallicity-indicating line ratios, such as N V/C IV, N V/He II, can be used to estimate metallicity by using the photoionization models  \citep{2002ApJ...564..592H,2006A&A...447..157N}. 

The measurements of the metallicity in quasar BLR have led to three intriguing findings: 1. Super solar metallicity in the BLR of luminous quasars at high redshift. \citet{2003ApJ...589..722D} found that the metallicity of quasar BLR is at least four times of the solar metallicity ($\rm Z_{\sun}$) in 70 most luminous $z\ge3.5$ quasars. These authors suggested intense star formation in quasar host galaxies at the early epoch of the Universe to enrich the quasar BLR. 2. There exists a strong correlation between the metallicity of quasar BLR and SMBH mass as well as quasar luminosity  \citep[e.g.,][]{2003ApJ...596...72W,2006A&A...447..157N,2011A&A...527A.100M}. 3. However, the metallicity of quasar BLR does not evolve with cosmic time  \citep{2003ApJ...596...72W, 2006A&A...447..157N}.

The evidence that there is no redshift evolution of the chemical abundance in the quasar BLR is quite puzzling. Studies have found a strong metallicity evolution in galaxies over cosmic time. High-redshift star-forming galaxies have lower gas-phase metallicity than their low-redshift counterparts for a given stellar mass \citep{2006ApJ...644..813E,2008A&A...488..463M,2013ApJ...771L..19Z,2014ApJ...780..122L,2014ApJ...795..165S,2014ApJ...792....3M,2015ApJ...799..138S,2015ApJ...808...25S, 2016ApJ...822...42O,2016ApJ...822..103G}. 
 
One interpretation of the non-evolution of metallicity at different redshifts is that the quasars in the above studies are biased to the most luminous quasars which are hosted by the most massive galaxies for a given redshift. Given the fact that the evolution of the galaxy mass-metallicity becomes weaker towards the high mass end \citep{2008A&A...488..463M,2013ApJ...771L..19Z, 2016ApJ...822...42O}, the metallicities in these luminous quasars may not evolve dramatically across cosmic time as well. Therefore, it is essential to study the chemical abundance in quasars with a broad range of luminosity, BH mass, and redshift. 

To solve the above issues, we select a sample of {$\sim$130,000 quasars in the redshift range of {$2.25\leq z \leq 5.25$} from the Sloan Digital Sky Survey (SDSS) Data Release 12 \citep[DR12]{2015ApJS..219...12A}}. {The BH mass range of this sample is $\rm 10^{7.5} \leq M_{\rm BH} \leq 10^{10} M_{\sun}$ and the bolometric luminosity range is $\rm 10^{44.2} erg/s\leq L_{\rm bol} \leq 10^{48} erg/s$.} We divide this quasar sample into different redshift and BH mass (bolometric luminosity) bins and generate the corresponding composite spectra in order to investigate the relationship between metallicity and BH mass (bolometric luminosity) at different redshifts. This sample is the largest quasar sample used to study the mass (bolometric luminosity) - metallicity relation. With this large sample and the composite spectra with high signal-to-noise ratio (S/N), we can get a precise measurement of line ratio to estimate the metallicity of quasar BLR and then get a better statistical investigation for the evolution of metallicity in a wide range of BH mass and bolometric luminosity. {We mainly focus on the metallicity of BLR in this work instead of narrow line region (NLR) which is supposed to trace the spatial scale \citep{2006A&A...456..953B,2006A&A...459...55B} and the enrichment history of quasar host galaxies \citep{2010IAUS..267...73N}. This is because that some typical NLR metallicity indicators, like  [N II]/H$\alpha$ \citep{2012ApJ...756...51L,2014MNRAS.438.2828D}, has lines that exceed the wavelength coverage of SDSS spectra, especially at higher redshift. Due to this problem, it is not easy to do sufficient statistical analysis on NLR with a wide redshift range comparing to BLR. The metallicity in quasar BLR can be affected by local starbursts at the center of quasar host galaxies \citep{2010IAUS..267...73N}, so it may not well present the global metallicity in host galaxies. But comparing the metallicity in the quasar BLR and its host galaxies can provide crucial information on the different history of star formation and metallicity enrichment in different parts of the galaxies.}

This paper is organized as follows: In \S  \ref{sec:Sample and composite spectra}, we describe how we select quasars from the SDSS data and generate composite spectra in each BH mass (bolometric luminosity) and redshift bins. In \S \ref{sec:Metallicity measurement}, we describe the fitting method of broad emission lines, the metallicity measurement and the corresponding results. We compare the metallicity in the quasar BLRs with that in the quasar host galaxies and present some discussion and possible explanations on the discrepancy between them in \S \ref{sec:Discussion}. We summarize our main conclusions in \S \ref{sec:Conclusion}. We adopt ($\rm \Omega_{\rm tot}$, $\rm \Omega_{\rm M}$, $\rm \Omega_{\Lambda}$) = (1.0, 0.3, 0.7), $\rm H_{0}$ = 70 km $\rm s^{-1}Mpc^{-1}$ \citep{2007ApJS..170..377S} and solar oxygen abundance of 12 + log(O/H) = 8.69 \citep{2009ARA&A..47..481A} in this paper.

\section{Sample and composite spectra}
\label{sec:Sample and composite spectra}
\subsection{Sample selection}
{We select quasars from the SDSS DR12 quasar catalog \citep{2017A&A...597A..79P}. There are 297,301 quasars in the SDSS DR12 quasar catalog. Most of the quasars in SDSS DR12 are observed as part of SDSS/BOSS quasar survey. The detail of the target selection can be found in \citet{2012ApJS..199....3R}. We adopt redshift from visual inspection (Z$\_$VI) in SDSS DR12 quasar catalog and we refer to readers \citet{2017A&A...597A..79P} for details of their visual inspection process. 
We estimate the virial BH mass of DR12 quasars by using the calibration from \citet{2006ApJ...641..689V} (VP06). 
The equation is:
\begin{equation}
\label{BH mass eq}
\log(\frac{M_{\rm BH}}{M_{\sun}}) = 0.66+0.53\log(\frac{\lambda L_{1350{\angstrom}}}{\rm 10^{44}erg/s})+2\log(\frac{\rm FWHM_{\rm C IV}}{\rm kms^{-1}})
\end{equation}
The full width at half maximum (FWHM) of C IV emission line is from \citet{2017A&A...597A..79P} and we exclude all the quasars whose FWHM$\_$CIV equal to -1 which means C IV is not in the spectra. We define several line-free windows: 1350{\AA}$\sim$1360{\AA}, 1445{\AA}$\sim$1455{\AA}, 1700{\AA}$\sim$1705{\AA}, 1770{\AA}$\sim$1800{\AA},
2155{\AA}$\sim$2400{\AA}, 2480{\AA}$\sim$2675{\AA}, 1150{\AA}$\sim$1170{\AA}, 1275{\AA}$\sim$1290{\AA}, 2925{\AA}$\sim$3400{\AA}, to fit the power-law continuum and then estimate the monochromatic flux at 1350{\AA}. }

We estimate the bolometric luminosity of DR12 quasars by using the monochromatic flux at 1350{\AA} and a bolometric correction of 3.81 \citep{0067-0049-166-2-470}.

\begin{figure*}
\includegraphics[height=7.cm,width=9cm]{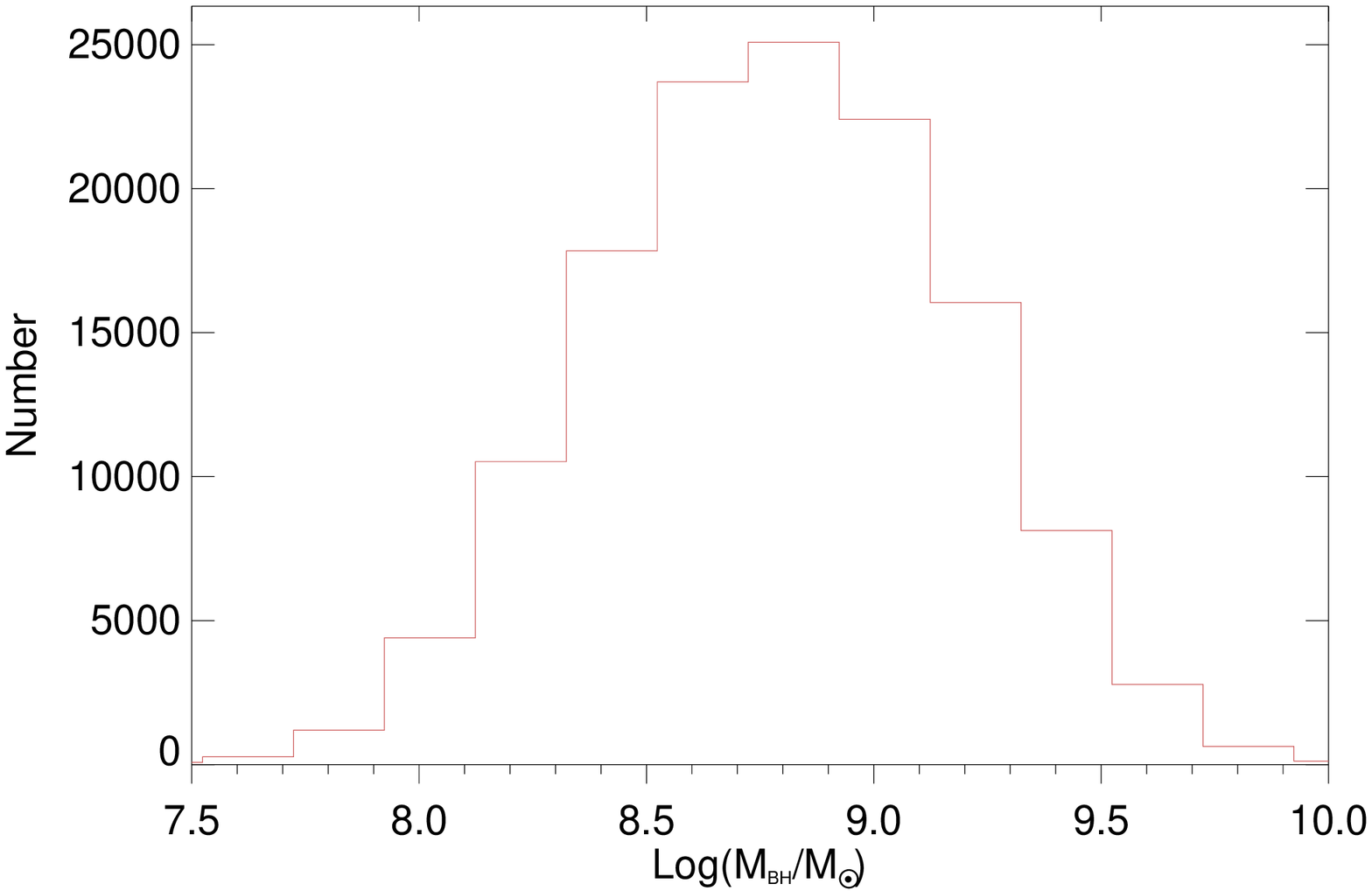}
\includegraphics[height=7.cm,width=9cm]{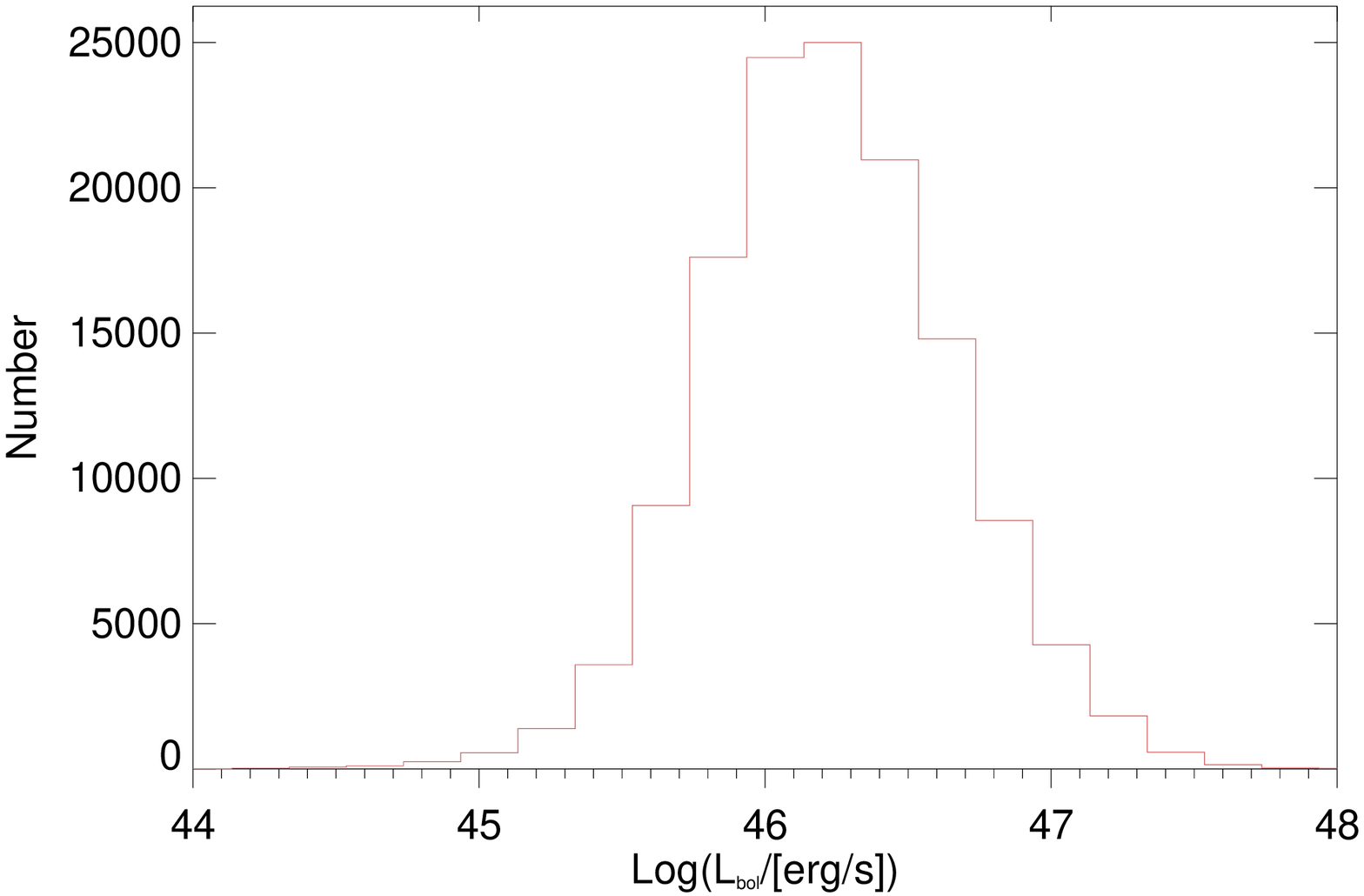}
\includegraphics[height=7cm,width=8.6cm]{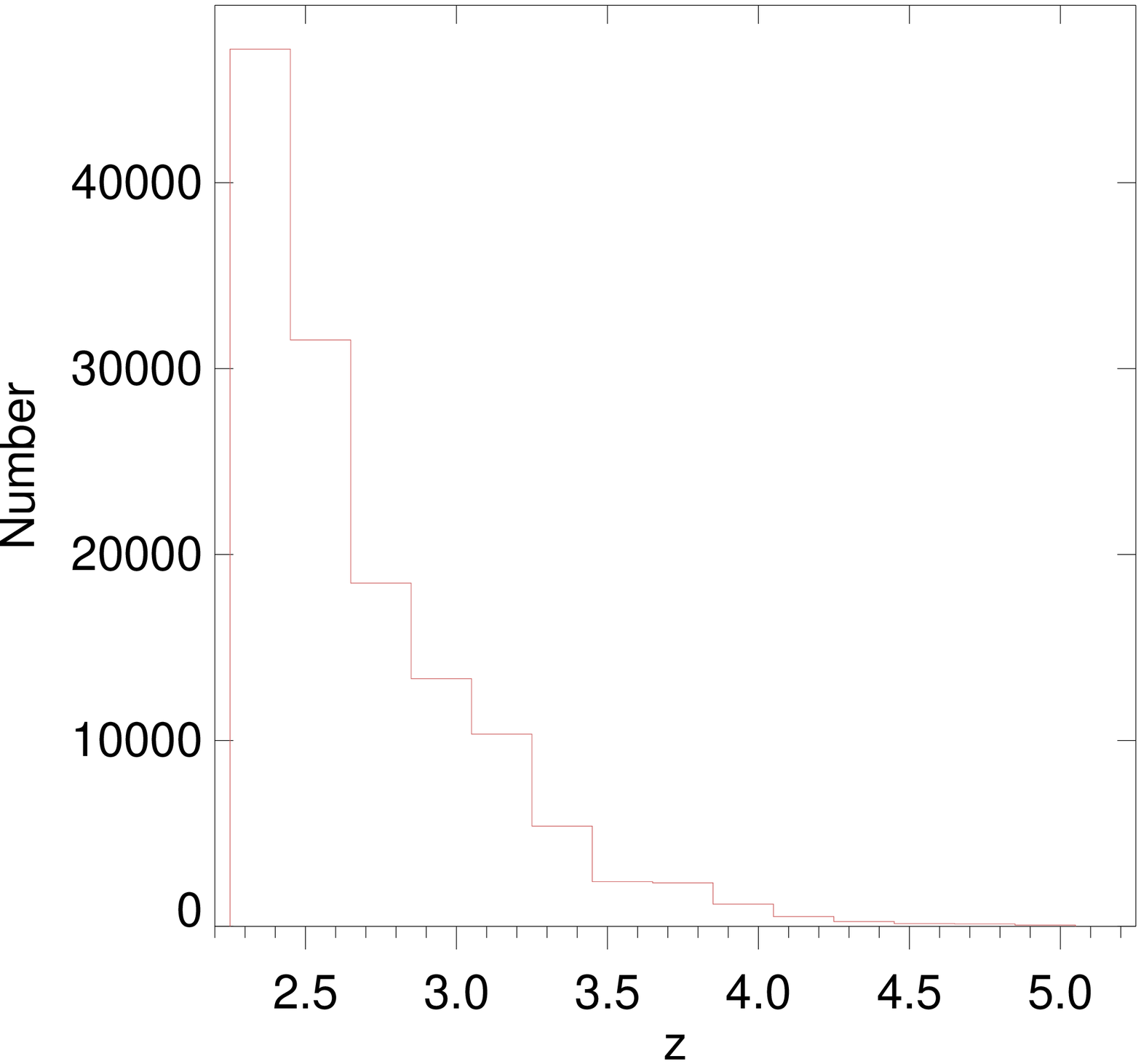}
\caption{{Histograms of our quasar sample selected from SDSS DR12. The redshift range is $2.25 < z < 5.25$ with the BH mass range $\rm 10^{7.5}M_{\sun} < M_{\rm BH} < 10^{10} M_{\sun}$, and bolometric luminosity range $\rm 10^{44.2} erg/s < L_{\rm bol} < 10^{48} erg/s$. Upper panel: The BH mass distribution of our sample. There are only 86 quasars in BH mass range $\rm 10^{6.5}M_{\sun} < M_{\rm BH} < 10^{7.5} M_{\sun}$ and 91 quasars in BH mass range $\rm 10^{10.0}M_{\sun} < M_{\rm BH} < 10^{10.6} M_{\sun}$ and we exclude them. Lower left panel: The bolometric luminosity distribution of our sample. The only 7 quasars in bolometric luminosity range $\rm 10^{48.0} erg/s < L_{\rm bol} < 10^{48.4} erg/s$ and 31 quasars in $\rm 10^{42} erg/s < L_{\rm bol} < 10^{44.2} erg/s$ are excluded. Lower right panel: The histogram of the redshift of our sample.}}
\label{fig:hist}
\end{figure*}

Then, we use the following criterion to select quasars: 
(i) {We exclude quasars with broad absorption lines (BAL) in the {SDSS DR12} quasar catalog using BAL$\_$FLAG$\_$VI, which were determined by visual inspection, {because BAL will affect the measurement on line flux and might also introduce uncertainty to the FWHM.} {{Quasars with FWHM$\_$CIV equal to -1 are also excluded. A total of 193,768 quasars are left after this step.}}
(ii) The redshift of selected quasars are between {2.25 and 5.25}.} Most UV metallicity diagnostic lines (e.g. N V, C IV) fall into SDSS spectroscopic wavelength coverage\footnote{The SDSS/BOSS wavelength coverage is from 3600{\AA} to 10500{\AA}.} within this redshift range. {{We only include quasars with ZWARNING = 0 which means that the pipeline redshift is very reliable. According to \citet{2017A&A...597A..79P}, the identification or redshift of 3.3$\%$ of these objects with ZWARNING = 0 changed after visual inspection. For the remaining 96.7$\%$, the redshifts from visual inspection are same with the pipeline redshifts. 17$\%$ of the overall SDSS sample have ZWARNING > 0 which means that the pipeline output may not be reliable. For these objects, only 41.1$\%$ of the identification or redshift changed after visual inspection. Others are still ZWARNING > 0. We exclude the quasars with ZWARNING > 0 in order to make sure the redshift measurement of our sample is secure enough. A total of 133,374 quasars are left after this step.}}
(iii) {We exclude quasars whose BH masses are not in the range of $\rm 10^{7.5} M_{\sun}\leq M_{\rm BH} \leq 10^{10.0} M_{\sun}$ and bolometric luminosity are not in the range of $\rm 10^{44.2} erg/s \leq L_{\rm bol} \leq 10^{48.0}erg/s$. Only a few quasars are out of this range and they unevenly scatter in a relatively wide BH mass or bolometric luminosity section, which makes it hard to divide them into bins, so we exclude them. Please see Figure \ref{fig:hist} for the details.} 

{Given these constrains on the SDSS DR12 quasar catalog, we select $\sim$130,000 quasars as our final sample to investigate the relationship between BH mass (bolometric luminosity) and quasar BLR metallicity at different redshifts.}

\subsection{Composite Spectra}
The main purpose of this work is to study the relationship between quasar BLR metallicity and BH mass (bolometric luminosity) at different redshifts. This requires us to measure the broad emission lines of quasar spectra in order to get the quasar BLR metallicity. Unfortunately, the individual SDSS spectra are too noisy to detect weak and highly blended metallicity diagnostic emission lines. Thus we study the metallicity in quasar BLRs by stacking quasar spectra in the same redshift and BH mass (bolometric luminosity) bins. 

We divide our sample into several redshift, BH mass and bolometric luminosity bins. Tables \ref{tab:samplegroupbh1}$\sim$\ref{tab:samplegroupbol2} summarize the redshift and BH mass (bolometric luminosity) bins and the number of quasars in each of the bins. Most bins have enough quasars for statistics and the sample is uniformly distributed in each bin as much as possible.


  \begin{table*}
 \centering
  \caption{We divide quasar sample within redshift $2.25<z<4.25$ from SDSS DR12 into 4 redshift bins and each redshift bin contains 11 BH mass bins. The number of the quasar in each bin is presented below. {We exclude some bins due to the poor S/N ratio of the composite spectra, and they are labeled with a `/'.}}
  \label{tab:samplegroupbh1}
  \begin{tabular}{cllll}
    \hline
Log($\rm M_{\rm BH}/M_{\sun})$ \textbackslash z & 2.25-2.75&2.75-3.25&3.25-3.75&3.75-4.25\\
\hline
7.5-7.8& 419 &119&/&/\\
7.8-8.0& 1462 &468&/&/\\
8.0-8.2&4584 &1649&227&/\\
8.2-8.4&9249 &3275&614&143\\
8.4-8.6& 14228 &4849&1037&309 \\
8.6-8.8& 17282 &5628&1442&403\\
8.8-9.0& 16838 &5473& 1608&470\\
9.0-9.2& 13493& 4707 &1695 &492 \\
9.2-9.4& 7729 &3276 & 1344 &458 \\
9.4-9.6& 2997  &1542 &707 &309\\
9.6-10.0&873  &555 &330 &125\\
    \hline
  \end{tabular}
 \end{table*}

 \begin{table*}
 \centering
  \caption{We divide the quasar sample within redshift range $4.25<z<5.25$ into 2 redshift bins and each redshift bin contains 2 BH mass bins. The number of the quasar in each bin is presented below.}
  \label{tab:samplegroupbh2}
  \begin{tabular}{cll}
    \hline
    \hline
Log($\rm M_{\rm BH}/M_{\sun})$  \textbackslash z& 4.25-4.75 & 4.75-5.25\\
\hline
8.00-9.25& 254 & 65 \\
9.25-10.00& 217 & 64 \\
    \hline
  \end{tabular}
 \end{table*}
 

 \begin{table*}
 \centering
  \caption{We divide quasar sample within redshift $2.25<z<4.25$ from SDSS DR12 into 4 redshift bins and each redshift bin contains 15 bolometric luminosity bins. The number of the quasar in each bin is presented below. {We exclude some bins due to the poor S/N ratio of the composite spectra, and they are labeled with a `/'. All the bins with bolometric luminosity $\rm 10^{44.2}erg/s-10^{44.6}erg/s$ are excluded due to the poor S/N of their composite spectra, so the real bolometric luminosity range in this study is $\rm 10^{44.6}erg/s-10^{48.0}erg/s$.}}
  \label{tab:samplegroupbol1}
  \begin{tabular}{cllll}
    \hline
    \hline
Log($\rm L_{\rm bol}/[erg/s]$)\textbackslash z & 2.25-2.75&2.75-3.25&3.25-3.75&3.75-4.25\\
\hline
44.6-45.0& 391 & / & / & /\\
45.0-45.2& 630 & 94 & / & /\\
45.2-45.5& 3144 & 521 & / & / \\
45.5-45.8& 12005 & 2353  & 217 & 37  \\
45.8-46.0& 15569 & 4216 & 479 & 77 \\
46.0-46.2& 17667 & 6122  & 1178 & 236 \\
46.2-46.4& 15585 & 6161 & 1862 &  435 \\
46.4-46.6& 11445 & 5053 & 202 & 660  \\
46.6-46.8& 6908 & 3529 & 1521 & 583 \\
46.8-47.0& 3541 & 1913 & 954 & 402 \\
47.0-47.2& 1550 & 1004 &  433 & 196  \\
47.2-47.4& 540 & 375 & 241 & 88 \\
47.4-47.6&  125 & 130 &  78 & 31 \\
47.6-48.0& 37& 38 & 24 & 14 \\

    \hline
  \end{tabular}
 \end{table*}


 \begin{table*}
 \centering
  \caption{We divide quasar sample within redshift $4.25<z<5.25$ from SDSS DR12 into 2 redshift bins and each redshift bin contains 3 bolometric luminosity bins. The number of the quasar in each bin is presented below.}
  \label{tab:samplegroupbol2}
  \begin{tabular}{cll}
    \hline
    \hline
Log($\rm L_{\rm bol}/[erg/s])$\textbackslash z  & 4.25-4.75 &4.75-5.25\\
\hline
 46.0-46.5 & 67 & 12 \\
 46.5-47.0 & 274 & 65 \\
 47.0-47.5 & 115 & 51 \\
    \hline
  \end{tabular}
 \end{table*}

 We generate the corresponding composite spectrum in each bin as follows:
\begin{enumerate}
   \item We obtain the reduced one-dimension quasar spectra from the SDSS DR12. 
  \item We correct Galactic extinction using the \citet{1998ApJ...500..525S} map and Milky Way extinction curve \citep{1989ApJ...345..245C}.
  \item We mask out 5-sigma outliers below the 20-pixel boxcar-smoothed spectrum to reduce the effects of narrow absorption \citep{2011ApJS..194...45S}. Bad pixels are also masked out.

  \item We shift each observed SDSS spectra into the rest frame based on the {visual redshift}.
  \item We re-sample each of the spectra with 1.0{\AA} bins into a unified wavelength range from 1000{\AA} to 2000{\AA} with a flux conservative manner for all redshift bins in range $2.25<z<3.75$. {The wavelength range is from 1000{\AA} to 1800{\AA} for quasars at $3.75<z<4.25$ and 1000{\AA} and 1600{\AA} for quasars at $4.25<z<4.75$ and $4.75<z<5.25$, due to the poor sensitivity beyond 10000$\angstrom$.} This rest-frame wavelength range covers all the lines used to estimate metallicity.
  
  \item {Because we are more interested in the emission line properties but not the continuum of the quasar spectra, we calculate the arithmetic mean flux instead of the geometric mean at each wavelength pixel to generate the composite spectra \citep{2001AJ....122..549V}.} We derive the errors of the composite spectra by using the Monte Carlo (MC) simulation \citep{2012ApJ...751...51J}. We first randomly choose $90\%$ of total individual spectra for a given BH mass (bolometric luminosity) and redshift bin and generate a `fake' composite spectrum by following the above steps from (i) $\sim$ (v). We repeat this for 1000 times and generate 1000 `fake' composite spectra. We then calculate the standard deviation of these `fake' composite spectra at each wavelength pixel and take them as the uncertainties of the real composite spectrum.
\end{enumerate}

These high S/N ($\sim$100 per resolution element) composite
spectra enable us to detect the weak emission lines to
measure the quasar BLR metallicity. {However, we exclude several composite sepctra due to their relatively poor S/N ($\sim$10 per resolution element). Please see Tables \ref{tab:samplegroupbh1}$\sim$\ref{tab:samplegroupbol2} for more details.}

\section{Metallicity measurement}
\label{sec:Metallicity measurement}
\subsection{Emission-line measurement}
\label{Method of emission-line measurement}
We fit the continuum of quasar composite spectra using a power-law function. We use the following line-free wavelength regions, $1445 \angstrom < \lambda < 1455 \angstrom$ and $1973 \angstrom < \lambda < 1983 \angstrom$, for the continuum fitting of composite spectra in the redshift bins $2.25-2.75, 2.75-3.25, 3.25-3.75$ \citep{2006A&A...447..157N,2011A&A...527A.100M}. We use wavelength regions $1445 \angstrom < \lambda < 1455 \angstrom$ and $1687 \angstrom < \lambda < 1697 \angstrom$ \citep{2006A&A...447..157N} for composite spectra in redshift bin $3.75-4.25$, and {$1275 \angstrom < \lambda < 1290 \angstrom$ and $1445 \angstrom < \lambda < 1455 \angstrom$ for composite spectra in redshift range $4.25-4.75$ and $4.75-5.25$ for power-law fitting because some line-free windows at longer wavelength have exceeded the coverage of SDSS spectrum at high redshift.}

Two methods are widely used to measure line flux \citep{2006A&A...447..157N}. One is by fitting the line with Gaussian function or Lorentzian function \citep{1997ApJ...475..469Z}. The other one is by integrating the line flux above a defined continuum flux without fitting the line profile \citep{2001AJ....122..549V}. Both approaches have limitation, especially when they are used to measure lines that are heavily blended with each other (e.g., Ly$\alpha$, N \uppercase\expandafter{\romannumeral5}). In this study, we adopt the method from \citet{2006A&A...447..157N} and \citet{2011A&A...527A.100M} to fit the emission line by the following function:
\begin{equation}
\label{eq:line}
\rm F_{\lambda}=\left\{
\begin{array}{rcl}
\rm F_{0}\times(\frac{\lambda}{\lambda_{0}})^{-\alpha}    &      & {\lambda > \lambda_{0}}\\
\rm F_{0}\times(\frac{\lambda}{\lambda_{0}})^{+\beta}     &      & {\lambda < \lambda_{0}}\\
\end{array} \right. 
\end{equation}
where we use 2 power-law indices, $\alpha$ and $\beta$, to control the red and blue sides of the emission line profile, respectively. $\lambda_{0}$ is the peak wavelength of the emission line, and $\rm F_{0}$ represents the peak intensity of the emission line. The ultraviolet lines are divided into two categories, high ionization lines (HILs), including N V, O IV, N IV], C IV and He II
and low ionization lines (LILs), including Si II, Si IV, O III], Al II,
Al III, Si III, and C III]
 \citep{1988PASP..100.1041C}. We use the same power-law indices for lines in the same category. When fitting Ly$\alpha$, we adopt the same $\alpha$ as that for HILs, but leave parameter $\beta$ free due to the intergalactic medium absorption of its blue wing \citep{2006A&A...447..157N}. 

\begin{figure*}
\includegraphics[width=3.2in]{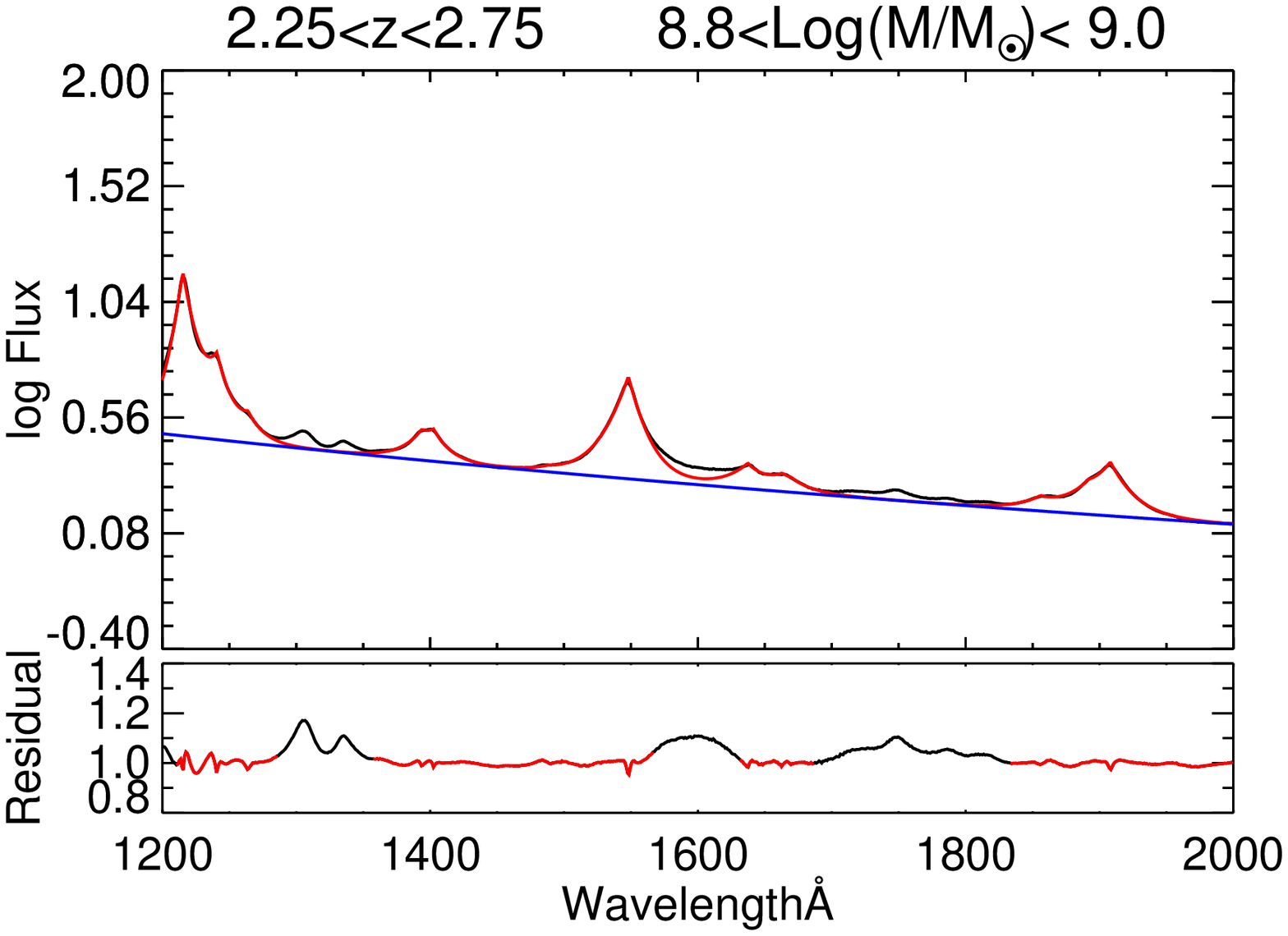}
\caption{An example of our composite spectra at redshift $2.25-2.75$ in BH mass bin $\rm 10^{8.8}M_{\sun}-10^{9.0}M_{\sun}$. The upper panel shows the original composite spectrum (black solid line) and the fitting spectrum (red solid line). The blue solid line indicates the power-law fitting of the quasar continuum. The lower panel shows the residual which is the ratio of the fitting spectrum and the composite spectrum. The regions included in the line-fitting process are shown in red and the regions excluded are shown in black in the lower panel. The unit of the flux density is $\rm 10^{-17}erg s^{-1}cm^{-2}\angstrom^{-1}$. }
\label{fig:fitting}
\end{figure*}

\begin{figure*}
\includegraphics[width=3.2in]{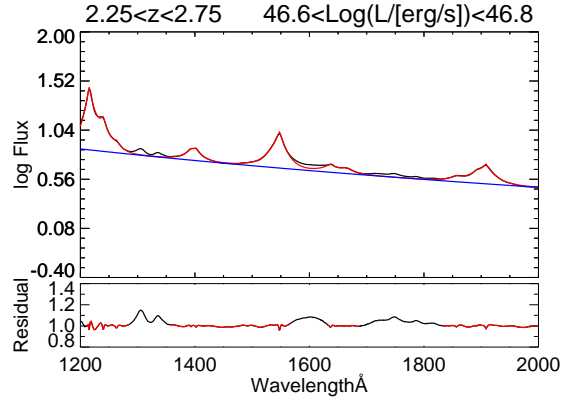}
\caption{An example of our composite spectrum at redshift $2.25-2.75$ in bolometric luminosity bin $\rm 10^{46.6}erg/s-10^{46.8}erg/s$. The upper panel shows the original composite spectrum (black solid line) and the fitting spectrum (red solid line). The blue solid line indicates the power-law fitting of the quasar continuum. The lower panel shows the residual which is the ratio of the fitting spectrum and the composite spectrum. The regions included in the line-fitting process are shown in red and the regions excluded are shown in black in the lower panel. The unit of the flux density is $\rm 10^{-17}erg s^{-1} cm^{-2}\angstrom^{-1}$. }
\label{fig:fittingbol}
\end{figure*}

We fit {Ly$\alpha$$\lambda$1216, N V$\lambda$1240, Si II$\lambda$1263, Si IV$\lambda$1398, O IV$\lambda$1402, N IV]$\lambda$1486, C IV$\lambda$1549, He II$\lambda$1640, O III]$\lambda$1663, Al II$\lambda$1671, Al III$\lambda$1857, Si III$\lambda$1887, C III]$\lambda$1909} simultaneously and allow the shift of the central wavelength, $\lambda_{0}$, with respect to the {rest-frame} wavelength within $\pm5${\AA} corresponding to about $\rm \pm1000km s^{-1}$. 
Unlike the {multi-Gaussian profile}, Function \ref{eq:line} only has four parameters for each line. When deblending lines that are close to each other, the fewer free parameters we use, the fewer ambiguities we will encounter. A more detailed discussion of this method and comparison between power-law profile and Gaussian function was presented in \citet{2006A&A...447..157N}. The fitting wavelength range for the emission lines is from $1210 \angstrom - 2000 \angstrom$ for {composite spectra in redshift 2.25 - 3.75, $1210 \angstrom - 1800 \angstrom$ for redshift 3.75-4.25, and $1210 \angstrom - 1600 \angstrom$ for redshift 4.25-5.25.} 
The undefined feature in $1570 \angstrom -1631 \angstrom$ which is called `1600{\AA} bump' is also excluded. The feature of 1600$\angstrom$ bump is ubiquitously presented in all our composite spectra and many previous study also noticed this feature \citep{1984MNRAS.207...73W,1990MNRAS.243..231B,1994ApJ...420..110L, 2006A&A...447..157N,2011A&A...527A.100M}. There is still a debate exists on the interpretation of it. \citet{2006A&A...447..157N} has given a very detailed discussion on the 1600$\angstrom$ bump. One possibility is that the 1600$\angstrom$ bump is one of the C IV component because a very redshifted broad component for Ly$\alpha$ and O VI was found \citep{1994ApJ...420..110L}. Another possibility is that the 1600$\angstrom$ bump is a blue shifted component of the He II emission because \citet{2006A&A...447..157N} found a similar negative correlation between the flux of 1600$\angstrom$/C IV with luminosity and He II/C IV with luminosity. The last possibility is that the 1600$\angstrom$ bump is caused by UV Fe II multiplet emission \citep{1994ApJ...420..110L}. Other heavy-blended lines such as O I+Si II composite, C II, N IV, Al II, N III], and Fe II multiplets are also excluded in our fitting. To sum up, the wavelength regions, $1286\angstrom-1357\angstrom$, $1570\angstrom-1631\angstrom$, and $1687\angstrom-1833\angstrom$, are excluded in our fitting process. Figures \ref{fig:fitting} and \ref{fig:fittingbol} are 2 typical examples of our fitting results. The upper panel shows the original composite spectrum, the best fitting spectrum and the continuum. The lower panel shows the residual which is the ratio between the observed spectrum and the best fitting spectrum. The residual is in the range of 0.9 $\sim$ 1.1 in most of the cases suggesting that our line fitting results well describe the line profiles in our composite spectra. 

As we have mentioned in the steps of generating composite spectra, we make 1000 `fake' composite spectra in the same BH mass (bolometric luminosity) and redshift bin and take their standard deviations as the uncertainties of the real composite spectra. Similarly, for the uncertainties of line ratios, we calculate 1000 line ratios of these 1000 `fake' composite spectra and derive the standard deviations as their corresponding errors. These uncertainties take both the variance of individual spectra {and systematics in the fitting procedure into account.} 

\subsection{Metallicity Measurements}
\label{sec:Line ratio to Metallicity}
Many studies have suggested that the broad emission-line flux ratios of quasars at the rest-frame ultraviolet and optical wavelength can provide accurate chemical abundance measurements in quasar BLRs \citep{2002ApJ...564..592H, 2003ApJ...596...72W, 2003ApJ...589..722D, 2006A&A...447..157N, 
2011A&A...527A.100M,2015Ap&SS.356..339M, 2017ApJ...834..203S}.  
The theoretical accuracy of any broad emission-line flux ratios as metallicity indicators depends on a variety of factors such as the central radiation field, temperature sensitivity of the emission-line ratio, the similarity of the ionization potentials and critical densities, the extent to which the line-emitting regions overlap spatially  \citep{2002ApJ...564..592H}. {In this work, we adopt 2 broad emission-line flux ratios to estimate the metallicity of quasar BLR: N \uppercase\expandafter{\romannumeral5}/C \uppercase\expandafter{\romannumeral4}, (Si  \uppercase\expandafter{\romannumeral4}+O  \uppercase\expandafter{\romannumeral4}])/C  \uppercase\expandafter{\romannumeral4}, which have been widely used in many previous studies \citep{1992ApJ...391L..53H,2002ApJ...564..592H,2003ApJ...589..722D,2006A&A...447..157N,2009A&A...494L..25J,2011A&A...527A.100M,2012ApJ...751L..23W}. N V, C IV, Si IV and O IV] are ubiquitously presented in all the quasar spectra at redshift 2.25$\sim$5.25. Using the same diagnostics for all the composite spectra will prevent us from introducing systematic error.} {N V/C IV can trace the chemical abundance base on the secondary nitrogen theory which suggests that N/H scales with $\rm Z^{2}$ (Z in this paper is O/H). (Si IV+O IV])/C IV, according to the simulation from \citet{2006A&A...447..157N}, shows significant correlation with BLR metallicity but not sensitive to the change of ionizing continuum. It traces the chemical abundance mainly because the relative importance of C IV as a coolant decreases as the BLR metallicity increases \citep{1996ApJ...461..683F,2009A&A...494L..25J,2012ApJ...751L..23W}.} We do not use line ratios relating N III], N IV], and Si II to estimate the metallicity because they are either highly blended with other strong lines or too weak to detect. {We apply these 2 broad emission-line flux ratios to all the redshift bins from 2.25 $\sim$ 5.25.}

{We transfer the above 2 broad emission-line flux ratios into metallicities by using the simulation results based on photoionization models from \citet{2002ApJ...564..592H} and \citet{2006A&A...447..157N}. The relation between the metallicity and broad emission-line flux ratio depends on the ionizing radiation field. \citet{2006A&A...447..157N} considered two possible SED models of the ionizing photons: one with a large UV thermal bump, the other has a weak UV thermal bump. \citet{2002ApJ...564..592H} considered three possible SED models which also span a wide range of possibilities from a strong `big blue bump' to a simple power law with no bump at all. We considered all the 5 SED models from \citet{2006A&A...447..157N} and \citet{2002ApJ...564..592H} for N V/C IV. \citet{2002ApJ...564..592H} didn't calculate the case for (Si IV+O IV])/C IV so we only considered 2 SED models from \citet{2006A&A...447..157N}. (Si IV+O IV])/C IV almost stays the same (no more than 0.05 dex change) with the change in SED (See Figure 29 in \citet{2006A&A...447..157N}), suggesting the metallicity estimated from (Si IV+O IV])/C IV is not very sensitive to SED.}

{Some of our emission-line flux ratios indicate a metallicity greater than 10 $Z_{\odot}$ which has exceeded the upper limit of the simulation from \citet{2002ApJ...564..592H} and \citet{2006A&A...447..157N}. {We assume that the relationship between emission-line flux ratio and the chemical abundance has the same trend when the metallicity is greater than 10 $Z_{\odot}$. We linearly extrapolate the points greater than 10 $Z_{\odot}$ in the log space. But it should be noticed that metallicity greater than 10 $Z_{\odot}$ is not calibrated and potentially unphysical.}}

{We calculate each metallicity by averaging the metallicity results from the considered different models. For the uncertainty \footnote{We noticed that this is not a real uncertainty but an estimation of the metallicity range.} of the metallicity, we take the highest metallicity derived from the considered SED models as the highest point of the error bar. The lowest points of the error bars are derived using the same way. Therefore, the metallicity errors in this paper take the systematic uncertainty introduced by different SED models into account} \footnote{The uncertainty of line measurement is usually very small, which is about 0.02 dex, comparing to the systematic error introduced by different SED models is about 0.5 dex. But in some special cases, like the first blue point in the right panel of Figure \ref{fig:2picbol}, where the error is extremely large, is due to poor fitting on emission line, not the uncertainty introduced by different SED models.}.

{In order to combine the metallicity results from different metallicity indicators, we average the metallicity derived from (Si IV+O IV])/C IV and N V/C IV. Tables \ref{tab:average metallicitybh1} $\sim$ \ref{tab:average metallicitybol2} present the final BLR metallicity in different redshift and BH mass (bolometric luminosity) bins along with error calculated from error propagation method.} 

\begin{table*}
 \centering
  \caption{The final metallicity ($\rm Z/Z_{\sun}$) of quasar BLR averaged from 2 different metallicity indicators in different BH mass bins at redshift 2.25-4.25.}
  \label{tab:average metallicitybh1}
  \begin{tabular}{cllll}
    \hline
    \hline
Log($\rm M_{\rm BH}/M_{\sun})$ \textbackslash z & 2.25-2.75&2.75-3.25&3.25-3.75&3.75-4.25\\
\hline
7.5-7.8& $3.74 ^{+ 0.62}_{- 0.80}$ &$3.59 ^{+ 0.63}_{- 0.74}$&/&/\\
7.8-8.0& $3.67 ^{+ 0.53}_{- 0.69}$ &$3.19 ^{+ 0.54}_{- 0.68}$&/&/\\
8.0-8.2& $3.72 ^{+ 0.49}_{- 0.65}$ &$3.29 ^{+ 0.48}_{- 0.64}$&$3.79 ^{+ 0.58}_{- 0.71}$&/\\
8.2-8.4&$3.92 ^{+ 0.51}_{- 0.67}$ &$3.61 ^{+ 0.52}_{- 0.68}$&$3.80 ^{+ 0.52}_{- 0.67}$&$4.18 ^{+ 0.67}_{- 0.83}$\\
8.4-8.6& $4.38 ^{+ 0.57}_{- 0.78}$ &$4.07 ^{+ 0.57}_{- 0.77}$&$4.37 ^{+ 0.61}_{- 0.82}$&$4.44 ^{+ 0.65}_{- 0.84}$ \\
8.6-8.8& $4.91 ^{+ 0.68}_{- 1.04}$ &$4.74 ^{+ 0.68}_{- 0.83}$&$5.24 ^{+ 0.83}_{- 0.96}$&$5.44 ^{+ 0.86}_{- 1.01}$\\
8.8-9.0& $5.91 ^{+ 0.88}_{- 1.04}$ &$5.91 ^{+ 0.97}_{- 1.14}$&$6.61 ^{+ 1.25}_{- 1.35}$&$6.99 ^{+ 1.30}_{- 1.45}$ \\
9.0-9.2& $7.43 ^{+ 1.34}_{- 1.52}$&$7.40 ^{+ 1.46}_{- 1.63}$&$8.43 ^{+ 1.71}_{- 1.83}$&$9.98 ^{+ 2.28}_{- 2.28}$ \\
9.2-9.4& $9.04 ^{+ 1.82}_{- 1.95}$ &$9.06 ^{+ 2.04}_{- 2.13}$ &$10.08 ^{+ 2.34}_{- 2.36}$&$11.32 ^{+ 2.85}_{- 2.75}$\\
9.4-9.6& $10.65 ^{+ 2.35}_{- 2.38}$  &$10.64 ^{+ 2.43}_{- 2.45}$ &$12.02 ^{+ 3.11}_{- 2.99}$ &$12.31 ^{+ 3.12}_{- 2.95}$\\
9.6-10.0& $13.03 ^{+ 3.26}_{- 3.11}$  &$13.46 ^{+ 3.50}_{- 3.32}$&$14.28 ^{+ 3.68}_{- 3.41}$&$20.19 ^{+ 5.54}_{- 4.84}$ \\
    \hline
  \end{tabular}
 \end{table*}

 
  \begin{table*}
 \centering
  \caption{The final metallicity ($\rm Z/Z_{\sun}$) of quasar BLR averaged from 2 different metallicity indicators in different BH mass bins at redshift 4.25-5.25.}
  \label{tab:average metallicitybh2}
  \begin{tabular}{cll}
    \hline
    \hline
Log($\rm M_{\rm BH}/M_{\sun})$ \textbackslash z& 4.25-4.75 & 4.75-5.25 \\
\hline
8.00-9.25& $3.66 ^{+ 0.60}_{- 0.74}$ & $5.47 ^{+ 1.39}_{- 1.28}$ \\
9.25-10.00 & $9.69 ^{+ 2.81}_{- 2.64}$ &$8.89 ^{+ 2.71}_{- 2.35}$ \\
    \hline
  \end{tabular}
 \end{table*}

 \begin{table*}
 \setlength{\tabcolsep}{6pt}
 \centering
  \caption{The final metallicity ($\rm Z/Z_{\sun}$) of quasar BLR averaged from 2 different metallicity indicators in different bolometric luminosity bins at redshift 2.25-4.25. All the bins with bolometric luminosity $\rm 10^{44.2}erg/s-10^{44.6}erg/s$ are excluded due to the poor S/N of their composite spectra, so the real bolometric luminosity range in this study is $\rm 10^{44.6}erg/s-10^{48.0}erg/s$.}
  \label{tab:average metallicitybol1}
  \begin{tabular}{cllll}
    \hline
    \hline
Log($\rm L_{\rm bol}/[erg/s]$)\textbackslash z & 2.25-2.75&2.75-3.25&3.25-3.75&3.75-4.25\\
\hline
44.6-45.0& $3.38 ^{+ 0.56}_{- 0.65}$ & / & / & /\\
45.0-45.2& $3.19 ^{+ 0.41}_{- 0.54}$ & $4.01 ^{+ 0.75}_{- 0.85}$ & / & /\\
45.2-45.5& $3.45 ^{+ 0.47}_{- 0.61}$ & $3.87 ^{+ 0.65}_{- 0.77}$ & / & / \\
45.5-45.8& $3.74 ^{+ 0.46}_{- 0.62}$ & $3.58 ^{+ 0.54}_{- 0.68}$  & $4.49 ^{+ 1.92}_{- 1.72}$ & $6.42 ^{+ 0.85}_{- 1.03}$  \\
45.8-46.0& $4.30 ^{+ 0.54}_{- 0.73}$ & $3.65 ^{+ 0.49}_{- 0.66}$ & $4.75 ^{+ 0.66}_{- 0.85}$ & $5.90 ^{+ 1.03}_{- 1.03}$ \\
46.0-46.2& $5.03 ^{+ 0.66}_{- 0.89}$ & $3.90 ^{+ 0.53}_{- 0.72}$ & $5.47 ^{+ 0.69}_{- 0.83}$ & $7.05 ^{+ 1.22}_{- 1.31}$ \\
46.2-46.4& $5.83 ^{+ 0.80}_{- 0.97}$ & $4.57 ^{+ 0.66}_{- 0.88}$ & $6.15 ^{+ 0.91}_{- 1.05}$ &  $6.95 ^{+ 1.23}_{- 1.33}$ \\
46.4-46.6& $6.84 ^{+ 1.13}_{- 1.28}$ & $5.27 ^{+ 0.80}_{- 0.96}$ & $7.08 ^{+ 1.26}_{- 1.38}$ & $7.65 ^{+ 1.36}_{- 1.50}$  \\
46.6-46.8& $7.86 ^{+ 1.50}_{- 1.66}$ & $6.03 ^{+ 1.04}_{- 1.20}$ & $8.26 ^{+ 1.63}_{- 1.76}$ & $8.28 ^{+ 1.51}_{- 1.63}$ \\
46.8-47.0& $8.94 ^{+ 2.02}_{- 2.09}$ & $6.64 ^{+ 1.28}_{- 1.44}$ & $8.96 ^{+ 2.02}_{- 2.06}$ & $10.41 ^{+ 2.55}_{- 2.50}$ \\
47.0-47.2& $10.22 ^{+ 2.49}_{- 2.51}$ & $7.83 ^{+ 1.82}_{- 1.89}$ & $10.02 ^{+ 2.45}_{- 2.45}$ & $11.40 ^{+ 3.00}_{- 2.85}$  \\
47.2-47.4& $11.50 ^{+ 3.01}_{- 2.94}$ & $7.76 ^{+ 1.85}_{- 1.86}$ & $11.38 ^{+ 3.10}_{- 2.95}$ & $11.36 ^{+ 2.88}_{- 2.74}$ \\
47.4-47.6& $13.02 ^{+ 3.86}_{- 3.53}$ & $8.55 ^{+ 2.19}_{- 2.15}$ & $12.06 ^{+ 3.53}_{- 3.27}$ & $15.93 ^{+ 6.13}_{- 4.64}$ \\
47.6-48.0& $15.51 ^{+ 4.77}_{- 4.16}$& $10.32 ^{+ 3.00}_{- 2.74}$ & $12.75 ^{+ 4.16}_{- 3.55}$ & $14.51 ^{+ 4.60}_{- 3.82}$ \\
    \hline
  \end{tabular}
 \end{table*}


 \begin{table*}
 \centering
  \caption{The final metallicity ($\rm Z/Z_{\sun}$) of quasar BLR averaged from 2 different metallicity indicators in different bolometric luminosity bins at redshift 4.25-5.25.}
  \label{tab:average metallicitybol2}
  \begin{tabular}{cll}
    \hline
    \hline
Log($\rm L_{\rm bol}/[erg/s])$\textbackslash z & 4.25-4.75 & 4.75-5.25 \\
\hline
46.0-46.5& $5.01 ^{+ 0.92}_{- 0.97}$ &$4.43 ^{+ 1.48}_{- 1.39}$ \\
46.5-47.0& $6.89 ^{+ 1.48}_{- 1.58}$&$8.38 ^{+ 2.53}_{- 2.33}$ \\
47.0-47.5& $8.87 ^{+ 2.48}_{- 2.30}$&$9.15 ^{+ 2.41}_{- 2.12}$\\
    \hline
  \end{tabular}
 \end{table*}

\begin{figure*}
\centering
\begin{minipage}[c]{0.5\textwidth}
\centering
\includegraphics[height=6.5cm,width=9cm]{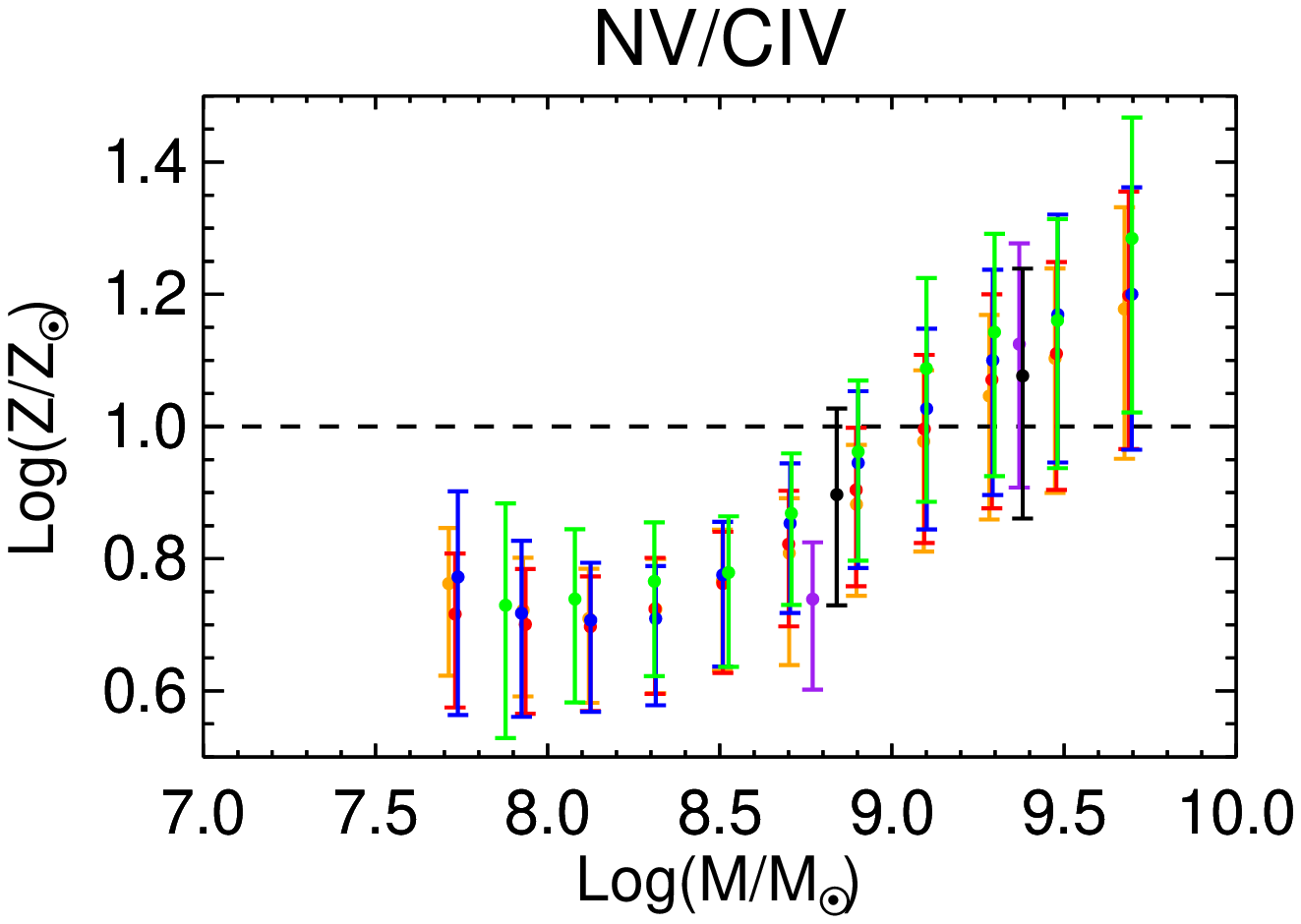}
\end{minipage}%
\begin{minipage}[c]{0.5\textwidth}
\centering
\includegraphics[height=6.5cm,width=9cm]{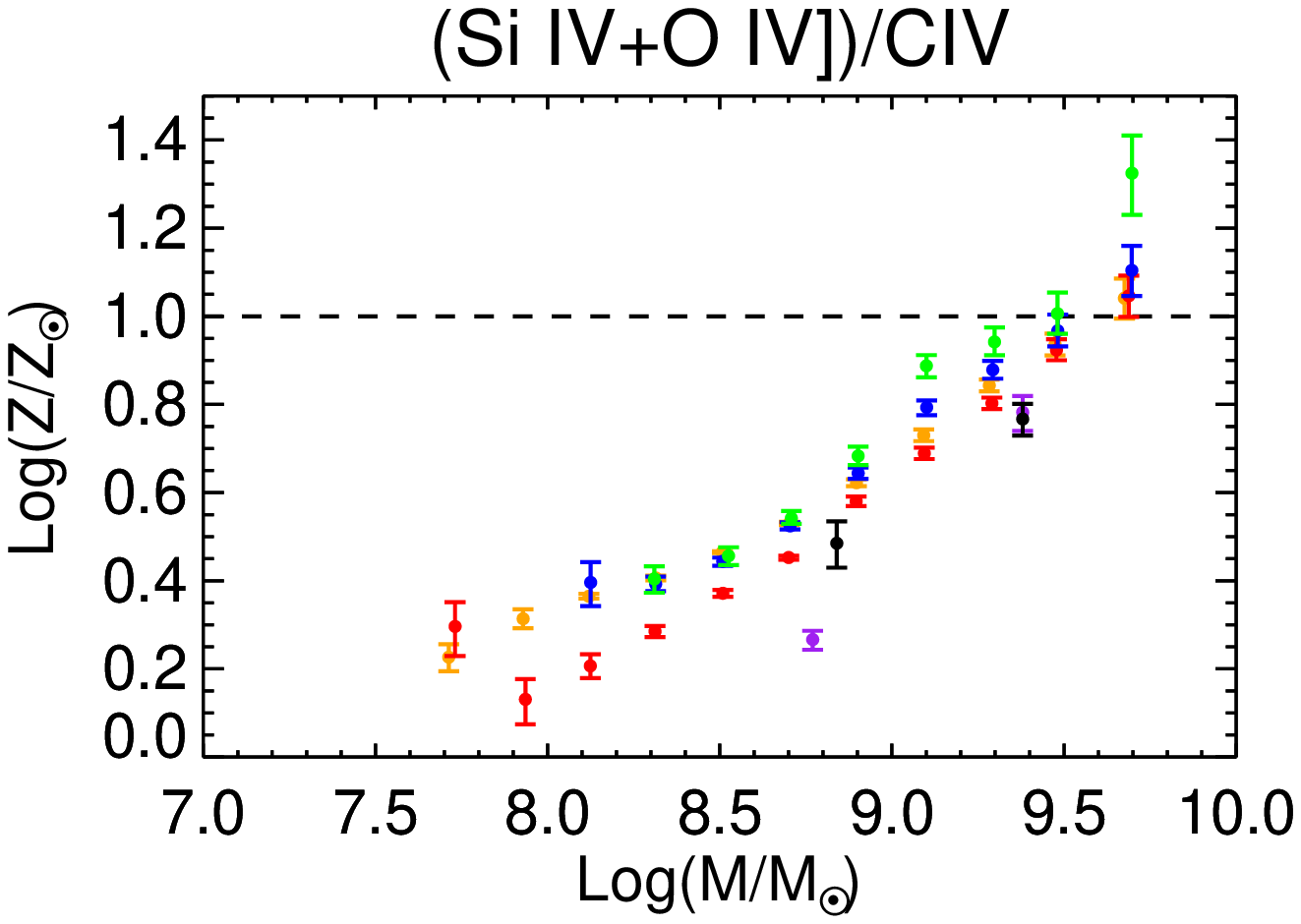}
\end{minipage}
\caption{Metallicity of quasar BLR versus BH mass. The metallicity of quasar BLR is derived from 2 different metallicity indicators,
which are labeled on top of each panel. Orange, red, blue, green, purple and black points represent the quasar BLR metallicity at redshift 2.25-2.75, 2.75
-3.25, 3.25-3.75, 3.75-4.25, 4.25-4.75, 4.75-5.25. The dash lines indicate 10 $Z_{\odot}$. Metallicity greater than 10 $Z_{\odot}$ is uncertain.}
\label{fig:2pic}
\end{figure*}

\begin{figure*}
\centering
\begin{minipage}[c]{0.5\textwidth}
\centering
\includegraphics[height=6.5cm,width=9cm]{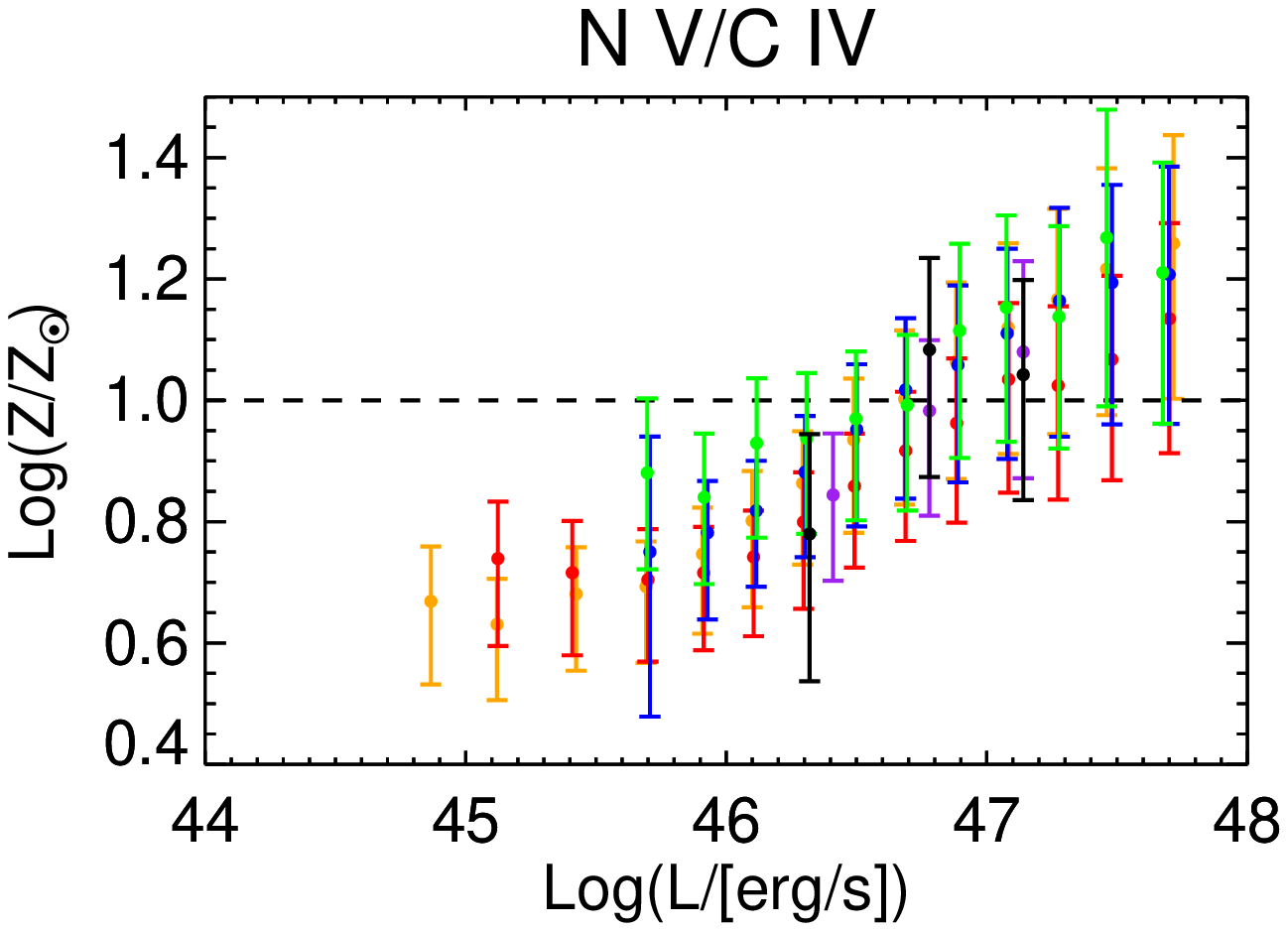}
\end{minipage}%
\begin{minipage}[c]{0.5\textwidth}
\centering
\includegraphics[height=6.5cm,width=9cm]{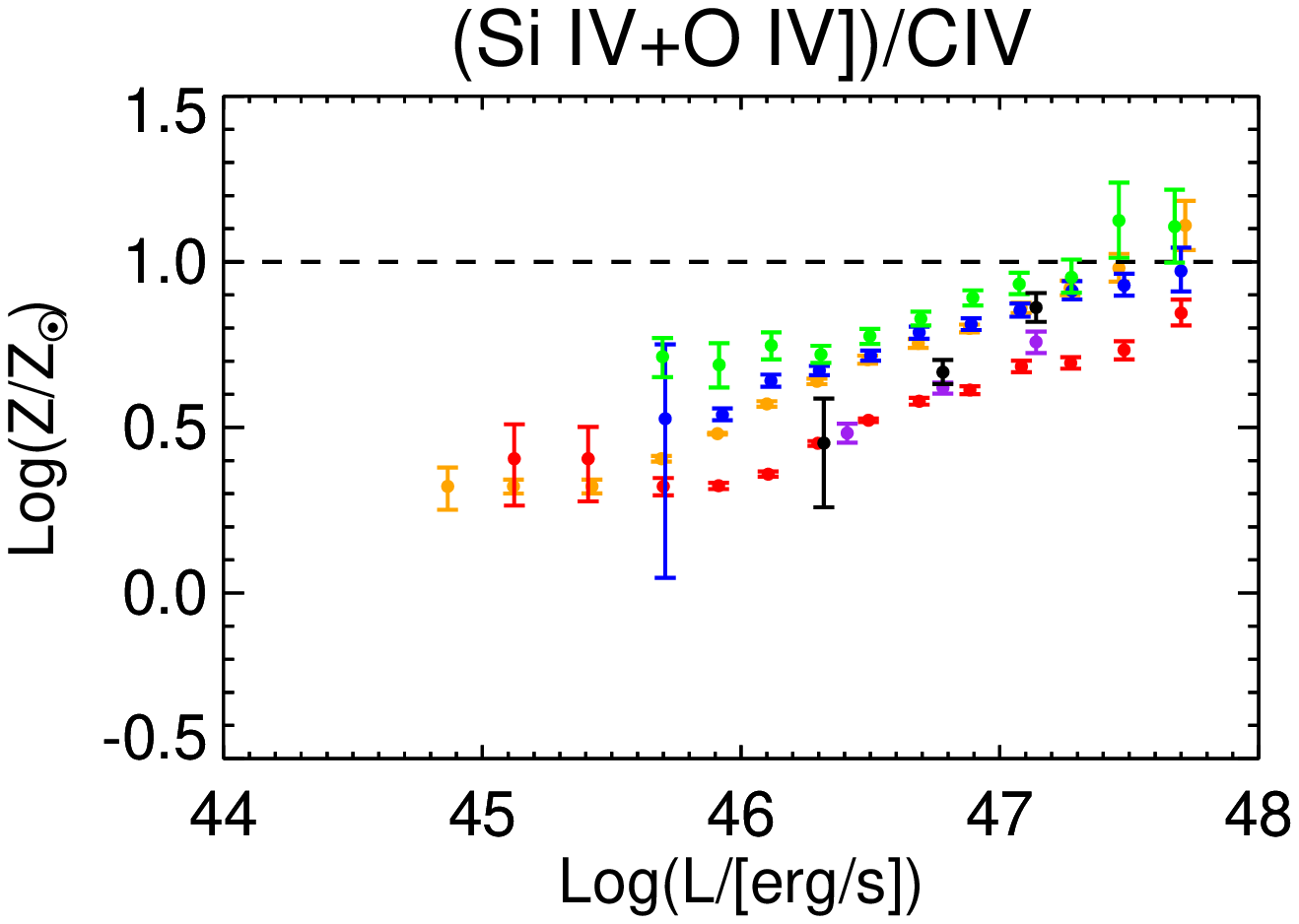}
\end{minipage}
\caption{Metallicity of quasar BLR versus bolometric luminosity. The metallicity of quasar BLR is derived from 2 different metallicity indicators,
which are labeled on top of each panel. Orange, red, blue, green, purple and black points represent the quasar BLR metallicity at redshift 2.25-2.75, 2.75
-3.25, 3.25-3.75, 3.75-4.25, 4.25-4.75, 4.75-5.25. The dash lines indicate 10 $Z_{\odot}$. Metallicity greater than 10 $Z_{\odot}$ is uncertain. {The lowest bolometric luminosity point in redshift range 3.25-3.75 of (Si IV+O IV])/C IV has a very large uncertainty due to the poor fitting of the emission line, not the systematic error from different SED models.}}
\label{fig:2picbol}
\end{figure*}

\begin{figure*}
\centering
\begin{minipage}[c]{0.5\textwidth}
\centering
\includegraphics[height=6.5cm,width=9cm]{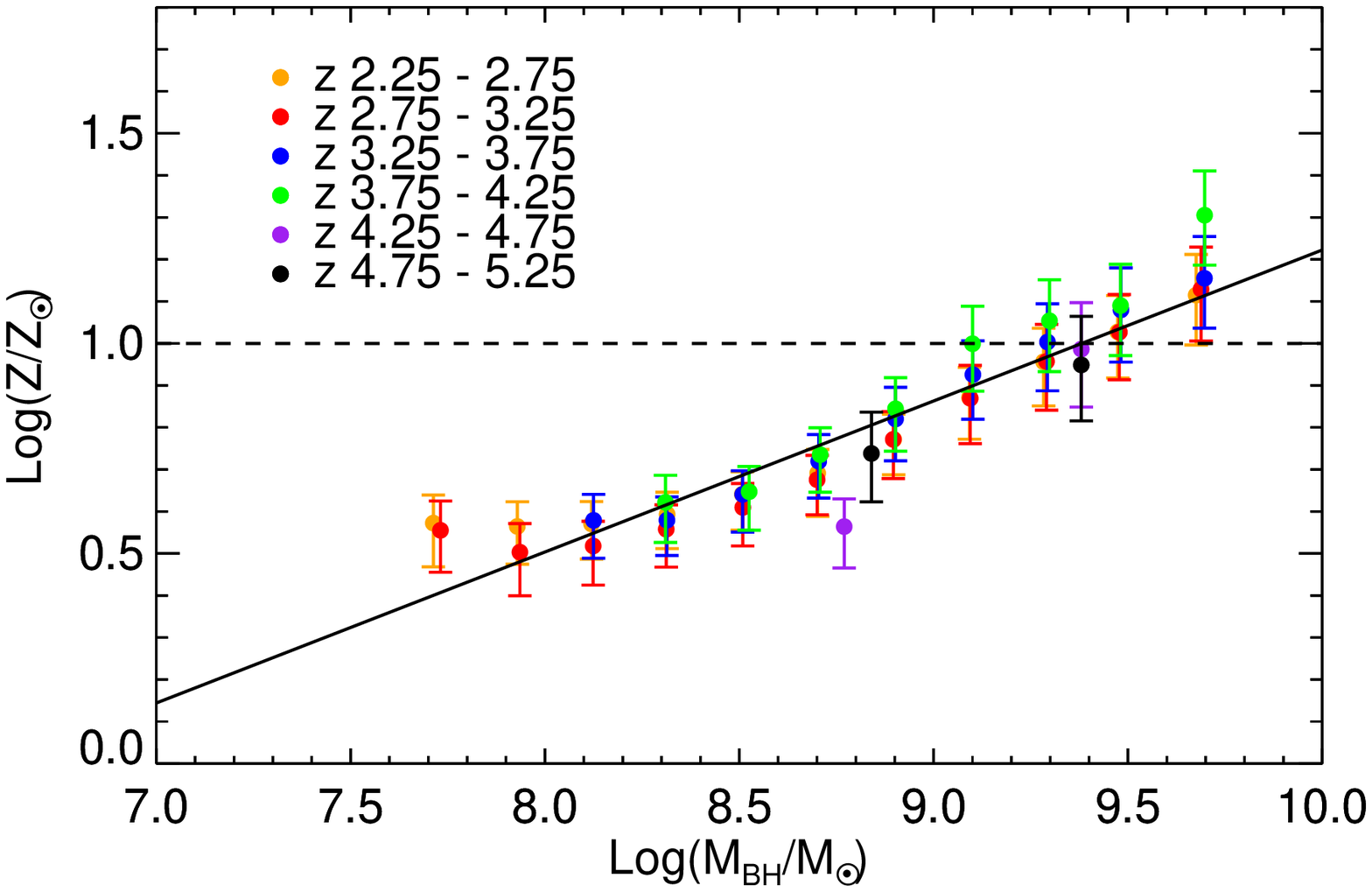}
\end{minipage}
\begin{minipage}[c]{0.5\textwidth}
\centering
\includegraphics[height=6.5cm,width=9cm]{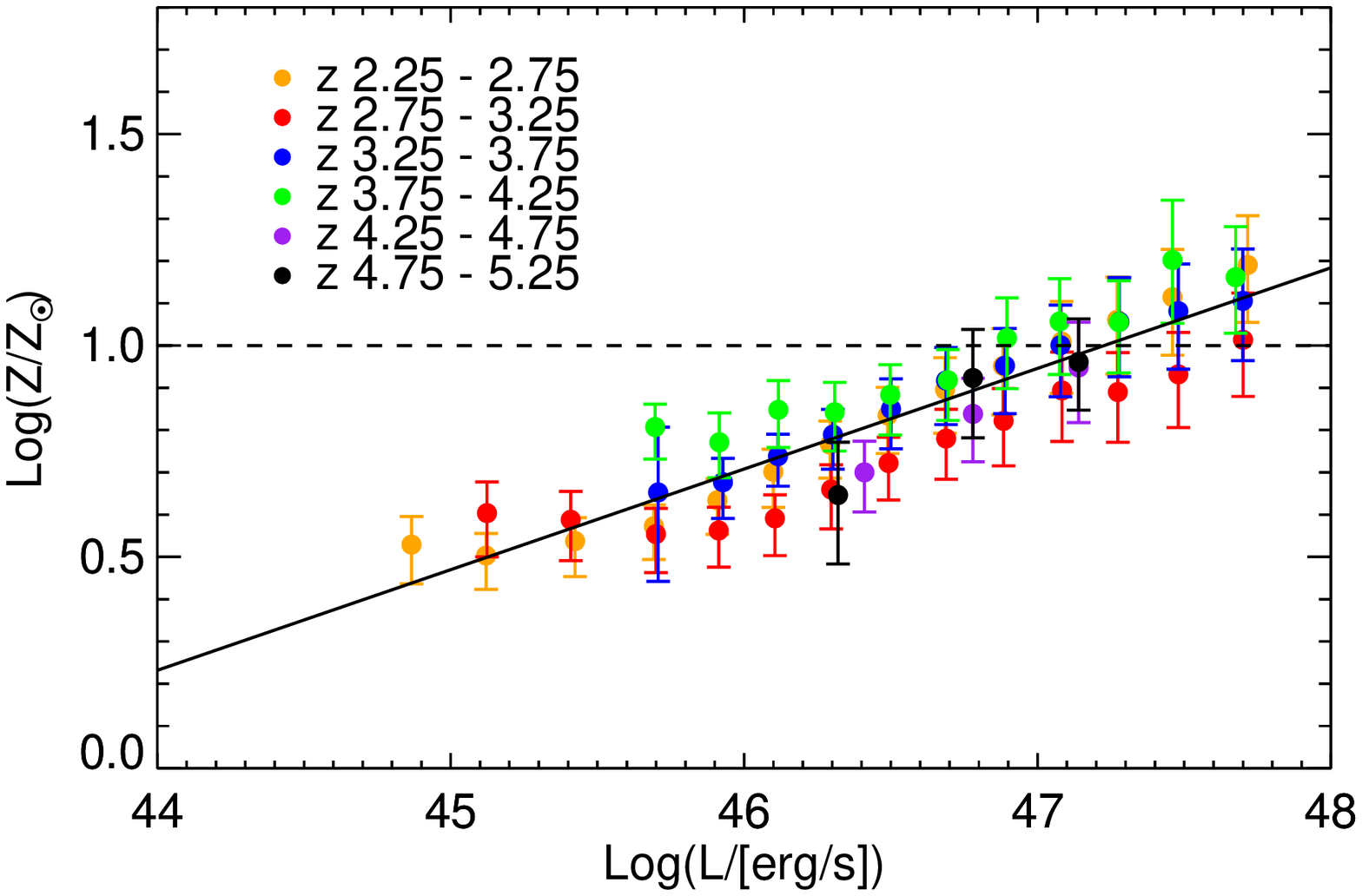}
\end{minipage}
\caption{Orange, red, blue, green, purple and black points represent the quasar BLR metallicity at redshift 2.25-2.75, 2.75
-3.25, 3.25-3.75, 3.75-4.25, 4.25-4.75, 4.75-5.25. Left panel: The relation between BH mass and the quasar BLR metallicity at different redshifts. We perform a
linear fit, y = kx + b, for this relation where $\rm b= -2.37 \pm 0.18$ and $\rm k = 0.36 \pm 0.02$. Right panel: The relation between bolometric luminosity and the quasar BLR metallicity at different redshifts. We perform a
linear fit, y = kx + b, for this relation where $\rm b= -10.24 \pm 0.68$ and $\rm k= 0.24 \pm 0.01$. Please see Tables \ref{tab:average metallicitybh1} $\sim$ \ref{tab:average metallicitybol2} for the
detailed metallicities and uncertainties shown on these 2 figures.}
\label{fig:Averpic}
\end{figure*}

\begin{figure*}
\includegraphics[width=5.5in]{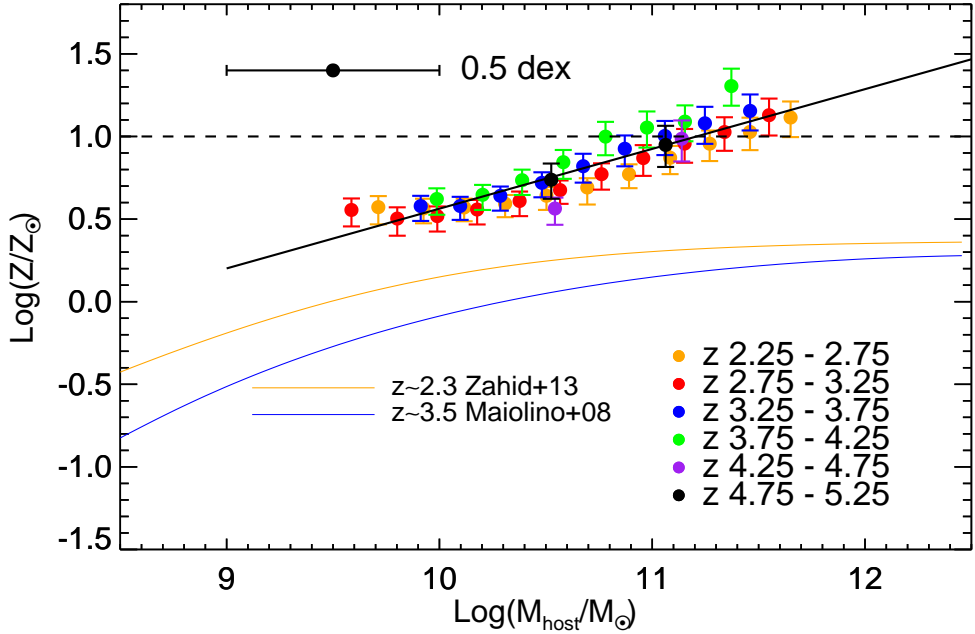}
\centering
\caption{Comparison of the metallicity in quasar BLRs with that in their host galaxies inferred from star-forming galaxies. Orange, red, blue, green, purple, and black points correspond to the quasar BLR metallicity at redshift $2.25-2.75, 2.75-3.25, 3.25-3.75$ ,$3.75-4.25$, $4.25-4.75$ and $4.75-5.25$ in different BH mass bins. {The error bar on the left upper corner of the figure shows the typical value of the uncertainty of the estimated host galaxies masses introduced by 1 $\sigma$ error of the fitting in \citet{2012MNRAS.420.3621T} on $\rm M_{BH}:M_{host galaxy}$.} As mentioned before, we also perform a linear fit (black solid line), $\rm y=kx+b$, for the correlation between quasar BLR metallicity and BH mass where $\rm b= -2.37 \pm 0.18$ and $\rm k = 0.36 \pm 0.02$. Orange solid line represents the mass-metallicity relationship of star-forming galaxies at $z\sim 2.3$ from \citet{2013ApJ...771L..19Z}. Blue solid line represents the mass-metallicity relationship of star-forming galaxies at $z\sim 3.5$ from \citet{2008A&A...488..463M}. The metallicity calibration of star-forming galaxies in this figure has all been transfered to KK04 calibration \citep{2004ApJ...617..240K}. The dash line indicates 10 $Z_{\odot}$. Metallicity greater than 10 $Z_{\odot}$ is uncertain. For the detailed value of the comparison result between quasar BLR and host galaxy, please see Tables \ref{tab:Discussion1} and \ref{tab:Discussion2}. }
\label{fig:Discussion}
\end{figure*}

 \begin{table*}
 \setlength{\tabcolsep}{2pt}
     \centering
  \caption{This table presents the masses of host galaxies transferred from BH masses, metallicities of quasar BLRs at redshift $2.25-2.75$, metallicities in corresponding host galaxies at $z\sim 2.3$ from \citet{2013ApJ...771L..19Z} in KK04 metallicity calibration and their differences {$\rm \Delta_{1} Log(\rm Z/Z_{\sun}$) (assuming $\rm M_{BH}:M_{host}$ evolves with redshift), $\rm \Delta_{2} Log(\rm Z/Z_{\sun}$) (assuming $\rm M_{BH}:M_{host}$ is constant) \citep{2012MNRAS.420.3621T}.}}
  \label{tab:Discussion1}
  \begin{tabular}{clllllllllll}
    \hline
    \hline
Log($\rm M_{\rm BH}/M_{\sun}$) Median &7.71&7.93&8.12&8.31&8.51&8.70&8.90&9.09&9.28&9.47&9.68\\
Log($\rm M_{\rm host}/M_{\sun}$)&$9.71 ^{+ 0.49}_{- 0.49}$ &$9.93 ^{+ 0.49}_{- 0.49 }$ &$10.12 ^{+ 0.49}_{- 0.49}$ &$10.31 ^{+ 0.49}_{- 0.49}$ &$10.50 ^{+ 0.49}_{- 0.49}$ &$10.69 ^{+ 0.49}_{- 0.49}$ &$10.89 ^{+ 0.49}_{- 0.49}$ &$11.08 ^{+ 0.49}_{- 0.49}$ &$11.27 ^{+ 0.49}_{- 0.49}$ &$11.46 ^{+ 0.49}_{- 0.49}$ &$11.65 ^{+ 0.50}_{- 0.50}$\\
Log($\rm Z_{\rm BLR}/Z_{\sun}$) &$0.57 ^{+ 0.07}_{- 0.10}$ &$0.56 ^{+ 0.06}_{- 0.09}$ &$0.57 ^{+ 0.05}_{- 0.08}$ &$0.59 ^{+ 0.05}_{- 0.08}$ &$0.64 ^{+ 0.05}_{- 0.09}$ &$0.69 ^{+ 0.06}_{- 0.10}$ &$0.77 ^{+ 0.06}_{- 0.08}$ &$0.87 ^{+ 0.07}_{- 0.10}$ &$0.96 ^{+ 0.08}_{- 0.11}$ &$1.03 ^{+ 0.09}_{- 0.11}$ &$1.11 ^{+ 0.10}_{- 0.12}$ \\
Log($\rm Z_{\rm host}/Z_{\sun}$) & $0.07 ^{+ 0.09}_{- 0.12}$ &$0.13 ^{+ 0.08}_{- 0.10}$ &$0.18 ^{+ 0.06}_{- 0.09}$ &$0.21 ^{+ 0.05}_{- 0.07}$ &$0.25 ^{+ 0.04}_{- 0.06}$ &$0.27 ^{+ 0.03}_{- 0.05}$ &$0.29 ^{+ 0.03}_{- 0.04}$ &$0.31 ^{+ 0.02}_{- 0.03}$ &$0.32 ^{+ 0.02}_{- 0.03}$ &$0.33 ^{+ 0.01}_{- 0.02}$ &$0.34 ^{+ 0.01}_{- 0.02}$ \\
$\rm \Delta_{1} Log(Z/Z_{\sun})$ &$0.50 ^{+ 0.18}_{- 0.20}$ &$0.43 ^{+ 0.16}_{- 0.17}$ &$0.39 ^{+ 0.14}_{- 0.15}$ &$0.38 ^{+ 0.13}_{- 0.14}$ &$0.39 ^{+ 0.12}_{- 0.13}$ &$0.42 ^{+ 0.11}_{- 0.14}$ &$0.48 ^{+ 0.10}_{- 0.11}$ &$0.56 ^{+ 0.11}_{- 0.12}$ &$0.63 ^{+ 0.11}_{- 0.12}$ &$0.69 ^{+ 0.11}_{- 0.12}$ &$0.77 ^{+ 0.11}_{- 0.13}$ \\
$\rm \Delta_{2} Log(Z/Z_{\sun})$ &$0.62 ^{+ 0.21}_{- 0.22}$ &$0.53 ^{+ 0.18}_{- 0.19}$ &$0.48 ^{+ 0.17}_{- 0.17}$ &$0.45 ^{+ 0.15}_{- 0.16}$ &$0.46 ^{+ 0.14}_{- 0.15}$ &$0.47 ^{+ 0.13}_{- 0.15}$ &$0.52 ^{+ 0.12}_{- 0.13}$ &$0.59 ^{+ 0.12}_{- 0.13}$ &$0.66 ^{+ 0.12}_{- 0.13}$ &$0.71 ^{+ 0.12}_{- 0.13}$ &$0.79 ^{+ 0.12}_{- 0.13}$ \\
    \hline
  \end{tabular}
 \end{table*}

    \begin{table*}
      \centering
      \setlength{\tabcolsep}{2pt}
  \caption{This table presents the masses of host galaxies transferred from BH masses, metallicities of quasar BLRs at redshift $3.25-3.75$, metallicities in corresponding host galaxies at $z\sim 3.5$ from \citet{2008A&A...488..463M} in KK04 metallicity calibration and their differences {$\rm \Delta_{1} Log(\rm Z/Z_{\sun}$) (assuming $\rm M_{BH}:M_{host}$ evolves with redshift), $\rm \Delta_{2} Log(\rm Z/Z_{\sun}$) (assuming $\rm M_{BH}:M_{host}$ is constant) \citep{2012MNRAS.420.3621T}.}}
  \label{tab:Discussion2}
  \begin{tabular}{clllllllllll}
    \hline
    \hline
Log($\rm M_{\rm BH}/M_{\sun}$) Median &8.13&8.31&8.51& 8.70&8.90&9.10&9.29&9.48&9.70\\
Log($\rm M_{\rm host}/M_{\sun}$) &$9.99 ^{+ 0.54}_{- 0.53}$ &$10.18  ^{+ 0.54}_{- 0.53}$ &$ 10.38 ^{+ 0.54}_{- 0.53}$ &$10.57 ^{+ 0.54}_{- 0.53}$ &$10.76 ^{+ 0.54}_{- 0.53}$ &$ 10.96 ^{+ 0.54}_{- 0.54}$ &$ 11.15 ^{+ 0.54}_{- 0.54}$ &$11.34 ^{+ 0.54}_{- 0.54}$ &$11.55 ^{+ 0.54}_{- 0.54}$\\
Log($\rm Z_{\rm BLR}/Z_{\sun}$)& $0.58 ^{+ 0.06}_{- 0.09}$ &$0.58 ^{+ 0.06}_{- 0.08}$ &$0.64 ^{+ 0.06}_{- 0.09}$ &$0.72 ^{+ 0.06}_{- 0.09}$ &$0.82 ^{+ 0.08}_{- 0.10}$ &$0.93 ^{+ 0.08}_{- 0.11}$ &$1.00 ^{+ 0.09}_{- 0.12}$ &$1.08 ^{+ 0.10}_{- 0.12}$ &$1.15 ^{+ 0.10}_{- 0.12}$ \\
Log($\rm Z_{\rm host}/Z_{\sun}$) &$-0.12 ^{+ 0.11}_{- 0.13}$ &$-0.06 ^{+ 0.09}_{- 0.12}$ &$-0.003 ^{+ 0.08}_{- 0.10}$ &$0.05 ^{+ 0.07}_{- 0.09}$ &$0.09 ^{+ 0.06}_{- 0.08}$ &$0.13 ^{+ 0.06}_{- 0.07}$ &$0.16 ^{+ 0.05}_{- 0.06}$ &$0.19 ^{+ 0.04}_{- 0.05}$ &$0.21 ^{+ 0.03}_{- 0.05}$ \\
$\rm \Delta_{1} Log(Z/Z_{\sun}$) & $0.69 ^{+ 0.19}_{- 0.20}$ &$0.64 ^{+ 0.17}_{- 0.18}$ &$0.64 ^{+ 0.16}_{- 0.17}$ &$0.67 ^{+ 0.16}_{- 0.16}$ &$0.73 ^{+ 0.16}_{- 0.16}$ &$0.80 ^{+ 0.15}_{- 0.16}$ &$0.84 ^{+ 0.15}_{- 0.16}$ &$0.89 ^{+ 0.15}_{- 0.17}$ &$0.94 ^{+ 0.15}_{- 0.15}$ \\
$\rm \Delta_{2} Log(Z/Z_{\sun}$) & $0.74 ^{+ 0.20}_{- 0.20}$ &$0.68 ^{+ 0.18}_{- 0.19}$ &$0.68 ^{+ 0.17}_{- 0.18}$ &$0.71 ^{+ 0.16}_{- 0.17}$ &$0.76 ^{+ 0.16}_{- 0.17}$ &$0.82 ^{+ 0.16}_{- 0.17}$ &$0.86 ^{+ 0.16}_{- 0.17}$ &$0.91 ^{+ 0.16}_{- 0.17}$ &$0.95 ^{+ 0.15}_{- 0.16}$ \\
    \hline
  \end{tabular}
 \end{table*}
    
\subsection{Results}
\label{sec:Result}
Figure \ref{fig:2pic} shows the relation between the metallicity indicated from the 2 broad emission-line flux ratios and BH mass at different redshifts. Figure \ref{fig:2picbol} shows the relation between the metallicity estimated from the 2 broad emission-line flux ratios and bolometric luminosity at different redshifts. 2 metallicity indicators all suggest that the metallicity of quasar BLR increases with BH mass and bolometric luminosity at all redshifts while no correlation between quasar BLR metallicity and redshift.

We examine the correlation between the metallicities derived from the above 2 broad emission-line flux ratios with BH mass, bolometric luminosity or redshift by adopting a Spearman rank-order test. Tables \ref{tab:Spearmantest table BH mass}, \ref{tab:Spearmantest table lbol} and \ref{tab:Spearmantest table redshift} present the Spearman rank-order correlation coefficients ($\rm r_{s}$) and the probability of the data being consistent with the null hypothesis that the metallicity is not correlated with BH mass, bolometric luminosity and redshift (p($\rm r_{s}$)), respectively. The larger the $\rm r_{s}$, the smaller the p($\rm r_{s}$), the more significant the correlation is. The Spearman rank-order test shows that there are significant positive correlations between the metallicities derived from (Si IV+O IV])/C IV, N V/C IV and BH mass. The Spearman rank-order test also shows that there is a significant correlation between the metallicity derived from (Si IV+O IV])/C IV, N V/C IV with bolometric luminosity. There is no significant correlation between metallicity and redshift presented in any 2 metallicity indicators.
 
In order to illustrate the trend between quasar BLR metallicity, BH mass (bolometric luminosity) and redshift, we thus have to average all the metallicities indicated from the 2 metallicity indicators as the final quasar BLR metallicity like we have mentioned before. The error of the final quasar BLR metallicity is calculated using error propagation. Figure \ref{fig:Averpic} shows the relationship of the final quasar BLR metallicity, BH mass (bolometric luminosity) and redshift. The metallicities of our quasar sample are super solar, ranges from {2.5 $\rm Z_{\sun}$ to 25.1 $\rm Z_{\sun}$}. They increase with BH mass and quasar luminosity. {As mentioned before, the metallicity greater than 10 $Z_{\odot}$ is uncertain. \citet{1984ApJ...285L..11F} also suggested that a high metallicity in photoionized gas clouds may result in a very low equilibrium temperature, and consequently a very low emissivity of emission lines. We give a dash line in Figures \ref{fig:2pic} $\sim$ \ref{fig:Discussion} to indicate the 10 $Z_{\odot}$. Values that exceed this line are highly uncertain.}

The Spearman rank-order test shows that there is statistically significant correlation between the final quasar BLR metallicity and BH mass while no correlation with redshift. There is also a siginificant correlation between quasar BLR metallicity and bolometric luminosity, though, the correlation, compared to that of to BH mass, is weaker. This result is also consistent with many former research \citep{2003ApJ...596...72W,2003ApJ...589..722D, 2006A&A...447..157N, 2011A&A...527A.100M}. {We also perform a linear fit, $\rm y=kx+b$, for the correlation between quasar BLR metallicity and BH mass, where $\rm b= -2.37 \pm 0.18$ and $\rm k = 0.36 \pm 0.02$ and the correlation between quasar BLR metallicity and bolometric luminosity, where $\rm b= -10.24 \pm 0.68$ and $\rm k= 0.24 \pm 0.01$ in Figure \ref{fig:Averpic}. }

   \begin{table}
  \centering
  \caption{The Spearman rank-order correlation coefficient $\rm r_{s}$ and the probability of the data being consistent with the null hypothesis that the metallicity is not correlated with BH mass, p($\rm r_{s}$). The last row presents the Spearman rank-order test results of the relationship between final quasar BLR metallicity averaged from the 2 metallicity indicators and BH mass.}
  \label{tab:Spearmantest table BH mass}
  \begin{tabular}{lcr}
    \hline
    \hline
Metallicity indicator & $\rm r_{s}$ & $ \rm p(r_{s})$\\
\hline
 (Si IV+O IV])/C IV&0.87&$7.5 \times$$10^{-15}$\\
 N V/C IV&0.95&$1.8 \times$$10^{-23}$\\
Average&0.97&$1.2 \times$$10^{-26}$\\
    \hline
  \end{tabular}
 \end{table}
  
   \begin{table}
   \centering
  \caption{The Spearman rank-order correlation coefficient $\rm r_{s}$ and the probability of the data being consistent with the null hypothesis that the metallicity is not correlated with bolometric luminosity, p($\rm r_{s}$). The last row presents the Spearman rank-order test results of the relationship between final quasar BLR metallicity averaged from the 2 metallicity indicators and bolometric luminosity.}
  \label{tab:Spearmantest table lbol}
  \begin{tabular}{lcr}
    \hline
    \hline
Metallicity indicator & $\rm r_{s}$ & $\rm p(r_{s})$\\
\hline
(Si IV+O IV])/C IV&0.83&2.8$\times$$10^{-15}$\\
N V/C IV&0.81&$8.1\times$$10^{-14}$\\
Average &   0.92 & 1.9$\times$$10^{-23}$  \\
    \hline
  \end{tabular}
 \end{table} 
  
   \begin{table}
   \centering
  \caption{The Spearman rank-order correlation coefficient $\rm r_{s}$ and the probability of the data being consistent with the null hypothesis that the metallicity is not correlated with redshift, p($\rm r_{s}$). The last row presents the Spearman rank-order test results of the relationship between final quasar BLR metallicity averaged from the 2 metallicity indicators and redshift.}
  \label{tab:Spearmantest table redshift}
  \begin{tabular}{lcr}
    \hline
    \hline
Metallicity indicator & $\rm r_{s}$ & $\rm p(r_{s})$\\
\hline
(Si IV+O IV])/C IV&0.42&4.6$\times$$10^{-3}$\\
N V/C IV&0.23 &0.14\\
Average &   0.16  & 0.31 \\
  \hline
  \end{tabular}
 \end{table}
 
\section{Discussion}
\label{sec:Discussion}
In this section, we compare the metallicities in quasar BLRs with those metallicities in their host galaxies. Due to the large brightness contrast between the quasar and their host galaxies, it is difficult to detect the rest-UV and/or optical metallicity diagnostic lines to measure their metallicity in high-redshift quasar host galaxies. In this study, we use the well-studied galaxy mass-metallicity relation \citep{2006ApJ...644..813E,2008A&A...488..463M,2013ApJ...771L..19Z} to infer the metallicity in quasar host galaxies.

First, we convert the BH mass ($\rm M_{\rm BH}$) to host galaxy mass ($\rm M_{\rm host}$) by adopting the evolution curve of $\rm M_{BH} : M_{host}$ ratio {from \citet{2012MNRAS.420.3621T}. These authors studied a sample of selected luminous ($\rm M_{\rm i}< 28$) SDSS quasar at $z\simeq 4$ using AO observations. They estimated the stellar masses of their host galaxies using the evolutionary synthesis models of \citet{2003MNRAS.344.1000B}, assuming the initial mass function (IMF) of \citet{1955ApJ...121..161S}, and using the C IV emission line to estimate the masses of their central SMBHs  \citep[see also,][]{0004-637X-824-2-70}. The curve in {Figure 6 of \citet{2012MNRAS.420.3621T}} has combined the observational results at different redshifts from \citet{2006MNRAS.368.1395M}, \citet{2006ApJ...649..616P}, and \citet{2003ApJ...587L..15W}.}
We find that the quasar host galaxies in this study are in the stellar mass range of $\rm 10^{9}M_{\odot}$ to $\rm 10^{12}M_{\odot}$, which spans a broad range of stellar mass. In order to infer the metallicities in the corresponding quasar host galaxies, we adopt the galaxy mass-metallicity relation at $z\sim2.3$ from \citet{2013ApJ...771L..19Z} and at $z\sim3.5$ from \citet{2008A&A...488..463M}. We convert the metallicities in \citet{2008A&A...488..463M} to \citet{2004ApJ...617..240K} metallicity calibration (KK04), which is the same as in \citet{2013ApJ...771L..19Z}.

In Figure~\ref{fig:Discussion}, we compare the metallicities in the quasar BLRs and the metallicity in their host galaxies. We plot the quasar BLR metallicity as a function of their host galaxy mass. For comparison, we also show the mass-metallicity relation of star-forming galaxies at $z\sim 2.3$ and $z\sim 3.5$. 
We find a significant discrepancy between the metallicity in quasar BLRs and their host galaxies. Tables \ref{tab:Discussion1} and \ref{tab:Discussion2} summarize the metallicity differences between the quasar BLRs and their host galaxies in different BH mass and redshift bins. 
We find the typical metallicities of the quasar BLRs is about {0.3-1.0 dex} higher than those in their host galaxies, when taking the uncertainty of the $\rm M_{\rm BH}/M_{\rm host}$ into account. This discrepancy is quite intriguing, considering that it is well believed that the gas that feeds the central SMBH is provided by their host galaxies. 

{We also notice that there is an active debate on whether the $\rm M_{BH}:M_{host}$ ratio evolves with redshift. Some studies suggest that the redshift evolution seen in $\rm M_{BH}:M_{host}$ ratio is probably caused by the selection effect and there is no intrinsic redshift evolution in this ratio \citep{2011A&A...535A..87S,2014MNRAS.438.3422S,2015ApJ...805...96S}. Therefore, we also convert the BH mass to host galaxy mass by adopting a constant ratio $\sim$0.01-0.05 from \citet{2012MNRAS.420.3621T}. This conversion is made by converting the BH mass using a ratio of 0.03 (midpoint) and then using 0.01 and 0.05 to calculate the upper and the lower limit of the estimated host galaxy mass. Tables \ref{tab:Discussion1} and \ref{tab:Discussion2} also summarize the metallicity discrepancy in this case. We found that the metallicity discrepancy is also about 0.3-1.0 dex which is approximately the same when we assume that $\rm M_{BH}:M_{host}$ ratio is evolved with redshift. Therefore, the result is robust against the choice of the two cases.}

We consider the following possibilities to explain the discrepancy of the metallicities in quasar BLRs and their host galaxies.

(1) The discrepancy between different metallicity diagnostic methods is a possible source to explain the discrepancy. There are three types of metallicity diagnostic methods: photoionization models, empirical calibrations, and direct-$\rm T_e$ methods  \citep{2008ApJ...681.1183K}. It has been long known that the discrepancy of metallicities estimated from different strong-line metallicity diagnostic methods can be as large as 0.5 dex, particularly between the photoionization models and direct-$\rm T_e$ method  \citep[e.g.,][]{2008ApJ...681.1183K, 2017ApJ...834...51B}. The systematical uncertainty of metallicity estimated from different photoionization models can be up to 0.2 dex \citep[e.g.,][]{2008ApJ...681.1183K, 2017ApJ...834...51B}. Considering that the differences of metallicity between the quasar BLRs and their hosts are {$0.3-0.8$ dex at $z=2.25-2.27$ and $0.4-1.0$ dex at $z=3.25-3.75$}, it is unlikely that the different photoionization models causes such large metallicity discrepancy between the quasar BLRs and their host galaxies. 

{{The uncertainty in the metallicity calibration for BLR can be a source causing the metallicity discrepancy. The oversimplification of the secondary nitrogen might affect the estimation of chemical abundance.} {H II region studies show that secondary N production will dominate when 12+log(O/H)>8.3 \citep{1976ApJ...204..330S,2000ApJ...541..660H}. It is obvious that the metallicity of quasar BLR has exceeded this value significantly, according to (Si IV+ OIV])/CIV which does not depend on the secondary nitrogen production theory. Studies on quasar BLR metallicity also show that the metallicity measured using nitrogen emission lines corresponds to that of using absorption line methods which is independent on the assumption of secondary N production \citep{1999ARA&A..37..487H,2002ApJ...564..592H}. Thus it is reasonable for us to believe that secondary N production is prominent in quasar BLR \citep{1999ARA&A..37..487H}. However, it is true that, based on some observational evidence \citep{1990ApJ...363..142G,1993A&A...277...42P, 1994MNRAS.270...35M,1999A&A...346..428P}, {there is a scatter in N/O with a fixed O/H caused by the bursts which temporarily lower N/O in the observed H II regions with sudden injections of fresh oxygen \citep{2000ApJ...541..660H}.} {\citet{1985ESOC...21..155P} suggests that there is an approximately $\pm$0.3 dex uncertainty in N/O at fixed O/H in the dwarf galaxies \citep{1990ApJ...363..142G}. We think that this cannot well explain the metallicity discrepancy between BLR and their hosts.} Local turbulence in BLR cloud could be another source of uncertainty. The internal Doppler velocity ($\rm v_{D}$) in BLR cloud is unknown \citep{2000ApJ...542..644B}. Calculation according to \citet{2002ApJ...564..592H} suggested that change of $\rm v_{D}$ is not significant to affect the line ratio related to nitrogen. We refer the readers to their paper for more details on the calculation. }}

(2) There is also the possibility that host galaxies of quasars do not follow the average mass-metallicity relation of galaxies. Galaxy metallicity depends not only on the stellar mass but also SFR. Studies show that host galaxies of luminous quasars are merger-triggered starburst galaxies and the SFR is about a few hundred to one thousandth $\rm M_{\sun}$ $\rm yr^{-1}$ \citep{2003A&A...406L..55B,2013ApJ...773...44W}. \citet{2011AJ....142..101W} observed nine $z\sim6$ quasars and found that the average SFR of the 5 millimeter-detected $\rm m_{1450}$ $\geq$ 20.2 quasars is about 560 $\rm M_{\sun}$ $\rm yr^{-1}$. \citet{0004-637X-824-2-70} studied 207 quasars selected from SDSS quasar catalogs and the Herschel Stripe 82 survey and find that their SFRs are about 500~$\rm M_{\sun}$$\rm yr^{-1}$. 
It is much larger than the SFR of UV-selected normal star-forming galaxies at $z\sim 2.3$ \citep{2006ApJ...644..813E} and $z\sim 3.5$ \citep{2008A&A...488..463M}, which are used to measure mass-metallicity relation. 
{According to the fundamental metallicity relation which states that the metallicity of galaxies with lower SFR is higher than that of galaxies with higher SFR for a given stellar mass \citep{2011MNRAS.414.1263M,2013MNRAS.434..451L}, the metallicity of quasar host galaxies should be less than or at least approximately equal to that of star-forming galaxies with the same stellar mass.}
This suggests that the metallicity of quasar host galaxy is probably overestimated based on the mass-metallicity relation of normal star-forming galaxies, which makes the discrepancy even larger. Therefore, the fundamental metallicity relation cannot explain the metallicity discrepancy between the quasar BLRs and quasar host galaxies.

(3) The metallicity gradient in quasar host galaxies is another possibility to cause this discrepancy. Studies show that luminous quasars exist at the late stage of the major  mergers \citep{2008ApJS..175..356H, 0004-637X-806-2-218}. \citet{0004-637X-753-1-5} studied the metallicity gradient in a sample of luminous infrared galaxies and found that the typical metallicity gradient in the late phase of the galaxy merger is about 0.02~dex/kpc. If considering the compact size of the quasar host galaxy  \citep[e.g.,$\sim1$~kpc,][]{2017arXiv170203852V}, the metallicity gradient of host galaxies is only no more than 0.05~dex. It could not explain such difference between the metallicity of quasar BLRs and host galaxies.

(4) {Another source of uncertainty is due to the BH mass estimation. This will affect our estimation on the stellar masses of quasar host galaxies.} The reliability of the C IV line to reproduce the more reliable H$\alpha$ or H$\beta$-based BH mass estimates is not well established \citep{2005MNRAS.356.1029B,2008ApJ...680..169S,2012ApJ...753..125S,2013MNRAS.434..848R}. The main criticism is that the single-epoch C IV profiles do not generally represent the reverberating BLR because of the existence of a low-velocity core component and a blue excess to the C IV emission, both of which do not reverberate \citep{2012ApJ...759...44D}. This is probably due to the contribution from an accretion disc wind \citep{2007ApJ...666..757S, 2011AJ....141..167R}, which results in a strong outflow or from a more distant narrow emission-line region \citep{1994ApJ...434..446K, 1995ApJ...451..498M, 2000ApJ...543..686P, 2005ApJ...631..689E, 2015MNRAS.451.2991G}. 
Due to the above skepticism, it is necessary for us to calibrate our BH mass. We calibrate our BH mass using 2 methods. The first method calibrates the C IV-based BH mass by using the trend of C IV-based BH mass/H$\alpha$-based BH mass versus C IV blueshift \citep{2016MNRAS.461..647C}. {The C IV blueshift (km $s^{-1}$) is defined as c $\times$ (1549.48{\AA}-fitted central wavelength of CIV in the composite spectra)/1549.48{\AA} and it is about 90$\sim$800 km $s^{-1}$ in this work. {We might underestimate the C IV blueshift because the $\rm Z\_VI$ is probably mostly determined by C IV or other luminous lines like Mg II. However, on average, the differences among $\rm Z\_VI$, $\rm Z\_Mg II$ and $\rm Z\_PCA$ (the redshift determined by using principal component analysis) are small which is <20 km $\rm s^{-1}$ \citep{2017A&A...597A..79P}. This will only bring $\sim$ 0.02 dex difference to the BH mass estimation which will not significantly affect our result.} The other method is calibrating the C IV-based BH mass with the peak ratio of $\lambda$1400 ({The definition of $\lambda$1400 is Si IV+O IV]}) and C IV \citep{2013MNRAS.434..848R}. {The first method shows that we probably underestimate the BH mass by $-0.2\sim 0.3$ dex while the second method shows $-0.3\sim 0.3$ dex which is quite consistent with the first one.} However, considering this uncertainty, the discrepancy especially in the higher BH mass bins still exists.} 

(5) It is worth noting that the possible change of quasar ionizing photon radiation field in different BH mass range that might also affect our results. In this work, we have considered different photoionization models covering a broad parameter space  \citep{2002ApJ...564..592H,2006A&A...447..157N}. In the most extreme case, we might underestimate the metallicity of the quasars in the lowest BH mass bin or overestimate the metallicity of the quasars at the highest BH mass bin by a factor of {about} 0.1 dex. However, even taking this effect into accounts, the mass-metallicity relationship still exists and it cannot explain the metallicity discrepancy.

(6){\citet{2004ApJ...608..108G} proposed that the fragmentation of quasar disk may result in supermassive star formation and the star will migrate inward to the central BH. \citet{2011ApJ...730...45J} carried out a further two-dimensional simulations. Star formation on quasar disk is a very interesting possibility to explain this metallicity discrepancy. {There is a few pieces of observational evidence in nearby quiescent galaxies, active Seyferts, and in our Galactic center \citep{2003ApJ...586L.127G,2005AJ....129.2138L,2007ApJ...671.1388D,2008ApJ...672L.119M} suggesting that stellar disks do appear within a few parsec or even $\sim 10^{-2}$pc of the SMBH, but a direct observation to resolve the star formation on quasar disk is still lacking \citep{2011ApJ...730...45J}.} The luminous supermassive star will blend in the light of quasar. \citet{2004ApJ...608..108G} and \citet{2011ApJ...730...45J} suggested that future periodicity searches or gravitational wave detections \citep{2003astro.ph..7084L} would bring more constrains on this topic. {There are also multiple theories suggesting different scenarios as well. \citet{2003astro.ph..7084L,2007MNRAS.374..515L} and \citet{1999Ap&SS.265..501C,1999A&A...344..433C} suggest that the fragmentation of the disks might results in the formation of many stars, even a nuclear stellar cluster, while \citet{2011ApJ...730...45J} suggest a single dominant mass. Besides the formation of the stars, whether the stars will eventually enrich the gas is another issue. A few theoretical studies show that the fragmentation of the unstable gaseous disk is able to give rise to the formation of protostars and consequently results in supernova explosion producing strong enriched outflows \citep{1999A&A...344..433C,2008A&A...477..419C,2011ApJ...739....3W}. Another possible scenario is that if the formed stars exceed a few hundred solar masses, the stars may disrupt themselves immediately upon reaching the zero-age main sequence due to the pulsational instabilities and then enrich CNO abundance of the surrounding diffuse gas by returning the mass after disruption \citep{2011ApJ...730...45J}. However, there is also possibility that the fragments may be accreted to the BH before contract to a single dominant mass \citep{2011ApJ...730...45J}. Also, as suggested by \citet{2004ApJ...608..108G}, the migration time of the stars might be comparable to their main-sequence life time so the stars might not be able to return their mass back to the gas by causing supernova-like explosion, resulting in no metallicity enrichment. We can not fully justify this possibility in the current paper, and it is also likely that the metallicity discrepancy is caused by a combination effect of (1) to (5).}}

\section{Conclusion}
\label{sec:Conclusion}
In this work, we used a large sample of quasar spectroscopic data from the SDSS DR12 with total $\sim$130,000 individual quasar spectra to investigate the metallicity of quasar BLR inferred from broad emission-line flux ratios based on photoionization models by fitting the composite spectra. The BH mass range and the bolometric luminosity range of the studied sample is $\rm 10^{7.5} M_{\sun}\leq M_{\rm BH} \leq 10^{10.0} M_{\sun}$ and $\rm 10^{44.6}erg/s\leq L_{\rm bol} \leq 10^{48.0} erg/s$ (Composite spectra in $\rm 10^{44.2}erg/s\leq L_{\rm bol} \leq 10^{44.6} erg/s$ are excluded due to the poor S/N.). Our main result can be summarized as follows:
\begin{enumerate}
\item The metallicity of quasar BLR ranges from {2.5 $\rm Z_{\sun}$ to 25.1 $\rm Z_{\sun}$} inferred from broad emission-line ratios (N \uppercase\expandafter{\romannumeral5}/C \uppercase\expandafter{\romannumeral4}, (Si \uppercase\expandafter{\romannumeral4}+O \uppercase\expandafter{\romannumeral4}])/C \uppercase\expandafter{\romannumeral4}). {Metallicity greater than 10 $Z_{\odot}$ is uncertain.} There is a statistically significant correlation between quasar BLR metallicity and BH mass (bolometric luminosity), but the metallicity does not evolve with redshift.
\item We compared the metallicity of quasar BLR with that of host galaxies inferred from the mass-metallicity relation of star-forming galaxy and find that the metallicity of quasar BLRs is higher than their host galaxies by {0.3 $\sim$ 1.0 dex}.

\item We considered several possibilities that cause the discrepancy, such as the effect of different SED models, systematic uncertainty of different metallicity diagnostic methods, mass-metallicity relations, C IV-based BH mass, the metallicity gradient in quasar hosts and some other possibilities. However, none of the above possibilities can well explain the large metallicity difference between the quasar BLRs and quasar host galaxies.

\item We proposed that the origin of the metallicity from quasar BLRs and their hosts may be different. Star formation probably occurs on quasar accretion disks which enriches the gas close to the BH and may causes this discrepancy. {However, there is no decisive observational evidence currently and the theory is also incomplete. Further studies are needed to justify this possibility.}
\end{enumerate}

\section*{Acknowledgements}
We thank Fred Hamann and the anonymous referee for comments that significantly improved the work, and Tohru Nagao for useful discussions. F.X. gratefully acknowledge the support from the undergraduate research program funding of Beijing Normal University. 
Y.S. acknowledges support from an Alfred P. Sloan Research Fellowship and NSF grant AST-1715579. Funding for the Sloan Digital Sky Survey IV has been provided by
the Alfred P. Sloan Foundation, the U.S. Department of Energy Office of
Science, and the Participating Institutions. SDSS-IV acknowledges
support and resources from the Center for High-Performance Computing at
the University of Utah. The SDSS web site is www.sdss.org.

SDSS-IV is managed by the Astrophysical Research Consortium for the 
Participating Institutions of the SDSS Collaboration including the 
Brazilian Participation Group, the Carnegie Institution for Science, 
Carnegie Mellon University, the Chilean Participation Group, the French Participation Group, Harvard-Smithsonian Center for Astrophysics, 
Instituto de Astrof\'isica de Canarias, The Johns Hopkins University, 
Kavli Institute for the Physics and Mathematics of the Universe (IPMU) / 
University of Tokyo, Lawrence Berkeley National Laboratory, 
Leibniz Institut f\"ur Astrophysik Potsdam (AIP),  
Max-Planck-Institut f\"ur Astronomie (MPIA Heidelberg), 
Max-Planck-Institut f\"ur Astrophysik (MPA Garching), 
Max-Planck-Institut f\"ur Extraterrestrische Physik (MPE), 
National Astronomical Observatories of China, New Mexico State University, 
New York University, University of Notre Dame, 
Observat\'ario Nacional / MCTI, The Ohio State University, 
Pennsylvania State University, Shanghai Astronomical Observatory, 
United Kingdom Participation Group,
Universidad Nacional Aut\'onoma de M\'exico, University of Arizona, 
University of Colorado Boulder, University of Oxford, University of Portsmouth, 
University of Utah, University of Virginia, University of Washington, University of Wisconsin, 
Vanderbilt University, and Yale University.



\bibliographystyle{mnras}
\bibliography{mybib} 





\bsp	
\label{lastpage}
\end{document}